\documentclass{aa}
\usepackage{newtxtext,newtxmath}
\usepackage{bm,enumitem}
\usepackage{url}
\usepackage{orcidlink} 
\usepackage{xcolor}
\definecolor{darkblue}{cmyk}{55,17,0,0}
\hypersetup{colorlinks=true,
  linkcolor=darkblue,
  linkbordercolor=darkblue,
  citecolor=darkblue,
  urlcolor=black
  }

\begin{document}

\title{Radio signatures of AGN-wind-driven shocks in elliptical galaxies}\subtitle{From simulations to observations}

\author{Haojie {Xia}\inst{1,2} \orcidlink{0009-0004-3881-674X}
    \and Feng {Yuan}\inst{3}\thanks{Corresponding author} \orcidlink{0000-0003-3564-6437}
    \and Zhiyuan {Li}\inst{4,5} \orcidlink{0000-0003-0355-6437}
    \and Bocheng {Zhu}\inst{6} \orcidlink{0000-0003-0900-4481}
}

\institute{Astrophysics Division, Shanghai Astronomical Observatory, Chinese Academy of Sciences, 80 Nandan Road, Shanghai 200030, P.R.China
    \and
    University of Chinese Academy of Sciences, No. 19A Yuquan Road, Beijing 100049, P.R.China
    \and
    Center for Astronomy and Astrophysics and Department of Physics, Fudan University, Shanghai 200438, P.R.China\\
    \email{fyuan@fudan.edu.cn}
    \and
    School of Astronomy and Space Science, Nanjing University, Nanjing 210023, P.R.China
    \and
    Key Laboratory of Modern Astronomy and Astrophysics, Nanjing University, Nanjing 210023, P.R.China
    \and
    National Astronomical Observatories, Chinese Academy of Sciences, 20A Datun Road, Beijing 100101, P.R.China}

\authorrunning{H. Xia et al.}
\titlerunning{Radio signatures of AGN-wind-driven shocks}

\date{Received / Accepted}

\abstract{We investigate the synchrotron emission signatures of shocks driven by active galactic nucleus (AGN) wind in elliptical galaxies based on our two-dimensional axisymmetric hydrodynamic MACER numerical simulations. Using these simulation data, we calculate the synchrotron radiation produced by nonthermal electrons accelerated at shocks, adopting reasonable assumptions for the magnetic field and relativistic electron distribution (derived from diffusive shock acceleration theory), and predict the resulting observational signatures. In our fiducial model, shocks driven by AGN winds produce synchrotron emission with luminosities of approximately $10^{29}\,\mathrm{erg\,s^{-1}\,Hz^{-1}}$ in the radio band (0.5-5 GHz), with spectral indices of $\alpha \approx -0.4$ to $-0.6$ during the strongest shock phases, gradually steepening to about $-0.8$ to $-1.4$ as the electron population ages. Spatially, the emission is initially concentrated in regions of strong shocks, later expanding into more extended, diffuse structures. We also apply our model to the dwarf elliptical galaxy Messier 32 (M32), and find remarkable consistency between our simulated emission and the observed nuclear radio source, suggesting that this radio component likely originates from hot-wind-driven shocks. Our results indicate that AGN winds not only influence galaxy gas dynamics through mechanical energy input but also yield direct observational evidence via nonthermal radiation. With the advent of next-generation radio facilities such as the FAST Core Array, SKA, and ngVLA, these emission signatures serve as important probes for detecting and characterizing AGN feedback.}

\maketitle
\section{Introduction}\label{sec:intro}
Over the past two decades, active galactic nucleus (AGN) feedback has emerged as a central ingredient in our understanding of galaxy formation and evolution. Both semi-analytical models and numerical simulations demonstrate that without AGN feedback, massive galaxies would be far more gas-rich and star-forming than observed, and the scaling relations between black holes and their host galaxies (e.g., the $M_\text{BH}-\sigma$ relation) would not naturally arise \citep{silk98, croton06,bower06,somerville15,yuan18}. Feedback from active galactic nuclei is now widely accepted as the dominant mechanism responsible for quenching star formation in massive systems, regulating the growth of supermassive black holes, and shaping the thermodynamic state of the interstellar medium (ISM) \citep{fabian12,kormendy13,heckman14,naab17}.

Active galactic nucleus feedback operates primarily in two distinct modes classified based on black hole accretion rate (BHAR). In the cold (or quasar) mode, which is associated with Eddington ratios higher than $2\%$, the strong wind launched by the luminous AGN can sweep out and/or heat the interstellar gas, and is believed to suppress star formation \citep[e.g.,][]{hopkins10,king15}. Observations of ultrafast outflows in X-ray spectra, with velocities up to approximately $0.1-0.3 c$, provide direct evidence for such winds in local AGNs \citep{tombesi10,gofford15}. On larger scales, outflows traced by optical, infrared, and molecular lines show that AGN-driven winds can reach kiloparsec distances and impact the host galaxy's ISM \citep{cicone14,harrison14,fluetsch19}.

In contrast, the hot (or kinetic, radio) mode dominates at lower accretion rates and is more prevalent in massive elliptical galaxies. Semi-analytic galaxy formation models demonstrated that such feedback is essential to halt cooling flows in massive halos and prevent the overproduction of bright galaxies \citep{croton06,bower06}. Cosmological simulations further showed that radio-mode feedback can quench star formation in cluster central galaxies and maintain hot gaseous halos \citep{sijacki07,dubois12}. Recent simulations revealed that the kinetic low-accretion mode plays a decisive role in controlling the black hole growth and the AGN luminosity \citep{weinberger17,yoon19,zhu23b} and in regulating the gas entropy within the galaxy and in the circumgalactic medium \citep{zinger20}. In this regime, magnetohydrodynamic (MHD) numerical simulations show that the AGN emits both jet and wind \citep{yuan12,yuan14,yuan15,yang21}. Although jets are more powerful than wind if the black hole spin is large, the momentum flux of wind is larger than that of the jet \citep{yang21}. In addition, the wind has a much larger opening angle and thus couples efficiently with the ambient ISM, making them potentially effective in depositing energy and momentum to the galaxy's gaseous halo \citep{dugan17,weinberger17}.

Despite their importance, hot winds are observationally more elusive than their cold-mode counterparts due to the technical difficulty of detecting them. Direct detections are achieved primarily through X-ray observations \citep{tombesi14,wang13,shi21,shi25}, while radio detections remain rare \citep{peng20}.

The observational evidence for wind-ISM interactions has emerged from multiwavelength studies. Most notably, the Fermi bubbles in the Galactic center are likely produced by the interaction of wind launched by the hot accretion flow in Sgr A* with the ISM \citep{mou14,mou15}. Additional supporting evidence was found recently in the X-ray band, where \cite{shi24} detected blueshifted emission lines in M81* and NGC 7213 that appear to originate from circumnuclear gas shock-heated by hot winds, providing strong observational evidence of wind-ISM interaction processes.

Previous studies have shown that wind-ISM interaction can produce shocks which accelerate electrons, potentially resulting in observable synchrotron emission, particularly in radio-quiet systems \citep{jiang10,nims15}. Such signatures offer a promising route to detect and characterize AGN winds even in systems where other tracers are weak or ambiguous.

However, these two works employed simplified AGN feedback models and were limited to one-dimensional simulations or calculations. The detailed spatial and spectral properties of such emission, as well as their detectability with current and future radio telescopes, have remained largely unexplored.

In this work, we investigate the synchrotron emission signatures of AGN-wind-driven shocks in elliptical galaxies using the simulation data obtained based on a model named MACER (Massive AGN Controlled Ellipticals Resolved) \citep{yuan18,yuan18b}. MACER is a numerical model that captures the evolution of a single galaxy with particular emphasis on AGN feedback effects, which is built on the ZEUS-MP code \citep{hayes06}, and based on over decades of work and development \citep[e.g.,][]{ciotti01,ciotti07,novak11,gan14}. Compared to earlier works, the main improvements in \cite{yuan18} is the incorporation of the state-of-the-art AGN physics, i.e., the properties of radiation and wind at various mass accretion rates. This improvement affected both the determination of the mass accretion rate at the black hole horizon and the feedback to the gas in the host galaxy. This model was later extended to high-angular-momentum elliptical galaxies \citep{yoon18}. We post-process the simulation data to identify shocks and model the acceleration of nonthermal electrons using the framework of diffusive shock acceleration. By adopting physically motivated assumptions for the magnetic field strength, we aim to compute the resulting synchrotron radiation across time and space and predict observational features that can serve as direct probes of AGN feedback in the radio band. We also apply our model to the nearby dwarf elliptical galaxy Messier 32 (M32) to test the viability of our approach against observational constraints. This study aims to provide theoretical guidance for interpreting current observations and designing future surveys with next-generation radio facilities such as the Five-hundred-meter Aperture Spherical radio Telescope (FAST) Core Array, Square Kilometre Array (SKA), and the next-generation Very Large Array (ngVLA).

The structure of the paper is as follows. In Section~\ref{sec:method}, we describe our numerical methods, including the hydrodynamic simulations, shock detection algorithm, and the calculation of nonthermal electron spectra and synchrotron emission. Section~\ref{sec:results} presents our main results, focusing on the temporal and spatial evolution of shock-driven synchrotron emission in our fiducial model of a massive elliptical galaxy. In Section~\ref{sec:result_m32}, we apply our model to the dwarf elliptical galaxy M32 and compare our predictions with recent radio observations. Section~\ref{sec:discussion} discusses the implications of our findings for understanding AGN feedback mechanisms and the observational prospects with next-generation radio facilities. Finally, we summarize our conclusions in Section~\ref{sec:conclusion}.

\section{Methods}\label{sec:method}
This study utilized hydrodynamic simulation data produced by the MACER model \citep{yuan18}. Our research primarily focused on the post-processing of these data, with particular emphasis on identifying shocks, computing shock-accelerated electron spectra, and tracking the evolution of nonthermal electron distributions.

We conducted two sets of simulations to explore the synchrotron emission signatures of AGN-wind-driven shocks. The first is a \texttt{fiducial model} representing a massive elliptical galaxy designed to investigate the general properties of AGN feedback in typical massive systems. The second is an \texttt{M32 model} specifically tailored to the nearby dwarf elliptical galaxy M32, enabling comparison with observations.

\subsection{The MACER simulation}\label{sec:hydro}
MACER is a two-dimensional numerical model for gas hydrodynamics that incorporates modules for star formation, stellar feedback, and AGN feedback. For the governing equations, we refer readers to \cite[][Eqs. 39-41]{yuan18}. AGN feedback is modeled using a subgrid black hole accretion framework that includes both cold and hot accretion modes, with the cold mode allowing for super-Eddington accretion. Jet feedback, however, is temporarily not included. One of the most important features of MACER is that the inner boundary of the simulation domain is smaller than the Bondi radius of the black hole accretion. In this case, we can calculate the inflow at the inner boundary. Combining with the black hole accretion theory, we can reliably obtain the mass accretion rate at the black hole horizon. The accretion rate determines the power of the AGN, which is the most important parameter for AGN feedback study. The stellar feedback module includes mass and energy injection from stellar winds as well as feedback from both Type Ia and Type II supernovae \citep{ciotti12}.

The computational domain is discretized using a grid in spherical coordinates $(r, \theta)$. The radial grid is logarithmically spaced, with the cell size $\Delta r$ kept smaller than 10\% of the local radius $r$ (i.e., $\Delta r/r \lesssim 0.1$), to achieve high spatial resolution near the galaxy center while covering a large radial extent. The angular grid is uniformly spaced. Both the inner ($r_\text{in}$) and outer ($r_\text{out}$) radial boundaries are set as outflow boundaries, allowing gas to leave the domain freely. The inner boundary, however, also serves as the injection site for AGN winds; when the feedback model dictates an outflow, the properties of the wind (density, velocity, and pressure) are imposed on the ghost zones. The polar axis is excluded by setting the angular domain to $[5^\circ, 175^\circ]$, with reflective boundary conditions applied at these boundaries. This is a numerical technique to avoid the numerical singularity at the pole and the associated severe timestep constraints. Since our current study focuses on wide-angle winds and does not include jets, this exclusion has a negligible effect on the results. For clarity and completeness, we briefly summarize the relevant modules implemented in MACER.

\subsubsection{AGN feedback}\label{sec:agn}
Based on whether the black hole accretion rate (BHAR) is above or below $2\% \dot{M}_\mathrm{Edd}$, the black hole accretion is categorized into two modes: cold and hot modes respectively \citep{yuan14}, where Eddington accretion rate is defined as
\begin{align}
    \dot{M}_\mathrm{Edd} = \frac{L_\mathrm{Edd}}{\epsilon_\text{EM}c^2},
\end{align}
and $L_\mathrm{Edd} = 4\pi G M_\mathrm{BH} m_\mathrm{H} c/\sigma_\mathrm{T}$ is the Eddington luminosity, $\epsilon_\text{EM}$ is the radiative efficiency usually set to 0.1 \citep{wu13}, $c$ is the speed of light, $M_\mathrm{BH}$ is the black hole mass, $G$ is the gravitational constant, $m_\mathrm{H}$ is the hydrogen mass, and $\sigma_\mathrm{T}$ is the Thomson scattering cross-section.

In the hot mode, the accretion flow exhibits a two-component structure: an inner hot accretion flow that transitions at a certain radius to an outer standard thin disk \citep{yuan14}. The truncation radius is approximated by \citep{yuan15}
\begin{align}
    r_\mathrm{tr}\approx 3 r_\mathrm{s} \left[\frac{0.02\,\dot{M}_\mathrm{Edd}}{\dot{M}(r_\mathrm{in})}\right]^2,
\end{align}
where $r_\mathrm{s}$ is the Schwarzschild radius, and $\dot{M}(r_\mathrm{in})$ is the inflow rate at the inner boundary. Strong winds are launched from the hot accretion flow, as we stated in the introduction. We adopt a subgrid model for the BHAR and wind properties that combines the results of \cite{yuan15} and \cite{cui20a, cui20b}. Specifically, \cite{yuan15} provides the wind properties based on general-relativistic magnetohydrodynamic (GRMHD) simulations of the accretion flow on small scales (e.g., $1000 r_g$), while \cite{cui20a, cui20b} extend these results by connecting the accretion flow scale to larger (e.g., $10^7 r_g$) scales through hydrodynamic and magnetohydrodynamic simulations. Our model thus incorporates the small-scale wind launching physics from GRMHD simulations and the large-scale propagation and interaction of winds as established by these bridging studies. The mass accretion rate at the black hole horizon is given by
\begin{align}
    \dot{M}_\mathrm{BH} \approx \dot{M}(r_\mathrm{in})\left(\frac{3r_\mathrm{s}}{r_\mathrm{tr}}\right)^{0.5}.
\end{align}
The wind properties are described by the mass outflow rate and velocity, which can be approximated as
\begin{align}
    \dot{M}_\mathrm{wind, hot} & \approx \dot{M}(r_\mathrm{in})-\dot{M}_\mathrm{BH}, \\
    v_\mathrm{wind, hot}       & \approx (0.2-0.4)v_\mathrm{K}(r_\mathrm{tr}),
\end{align}
where $v_\mathrm{K}(r_\mathrm{tr})$ represents the Keplerian velocity at the truncation radius. The polar angular distribution of the hot wind spans $\theta\sim 30^\circ-70^\circ$ and $110^\circ-150^\circ$. The radiative efficiency $\epsilon_\text{EM}$ of hot accretion flows follows the prescription of \cite{xie12}.

In the cold mode, when $\dot{M}_\text{BH} > 0.02\,\dot{M}_\mathrm{Edd}$, the accretion flow transitions to a standard thin disk extending to the innermost stable circular orbit (ISCO). The black hole's mass accretion rate is derived by solving a set of differential equations that describe the mass flow through its accretion disk \citep[][Eqs. 3-13]{yuan18}. This model considers the disk's primary mass source (infalling gas) and its two main channels of mass loss: accretion onto the black hole and outflows from disk winds. The wind properties in this mode are empirically constrained by observations of luminous AGNs, with mass outflow rates and velocities following the scaling relations with the luminosity of the AGN $L_\mathrm{BH}$ established by \cite{gofford15} where the radiative efficiency $\epsilon_\text{EM}$ in this regime is approximately $0.1$, yielding $L_\mathrm{BH}=0.1\,\dot{M}_\mathrm{BH}c^2$. Specifically, the mass outflow rate and wind velocity are given by
\begin{align}
    \dot{M}_\mathrm{wind,cold} & =0.28\left(\frac{L_\mathrm{BH}}{10^{45}\,\mathrm{erg\,s^{-1}}}\right)^{0.85}\,\mathrm{{M_\odot}\, yr^{-1}},                         \\
    v_\mathrm{wind,cold}       & =\min\left(2.5\times10^4\left(\frac{L_\mathrm{BH}}{10^{45}\,\mathrm{erg\, s^{-1}}}\right)^{0.4}, 10^5\right)\,\mathrm{km\, s^{-1}}.
\end{align}
Note that observations of cold winds indicate a velocity saturation at about $10^5\,\mathrm{km\,s^{-1}}$ \citep{gofford15}. The polar angular distribution of the cold wind is assumed to follow a $\cos^2\theta$ dependence.

The effect of AGN radiation is also incorporated through cooling (C) and heating (H) function from \cite{sazonov05} which model a plasma in photoionization equilibrium exposed to the average quasar spectral energy distribution. This function accounts for Compton heating and cooling, bremsstrahlung losses, photoionization heating, line and recombination continuum cooling. Assuming the flow is optically thin, the simplified radiative transfer equation which does not include the effects of dust can be integrated as
\begin{align}
    \frac{\mathrm{d}L_\text{BH}^\text{eff}(r)}{\mathrm{d}r} = -4\pi r^2 H,
\end{align}
where $L_\text{BH}^\text{eff}(r)$ is the effective AGN luminosity at radius $r$ and $H$ is the heating rate per unit volume due to AGN radiation. This equation is solved with a specific boundary condition $L_\text{BH}^\text{eff}(r=0)=L_\text{BH}$. The radiation pressure comes from photoionization, Compton opacity and electron scattering, whose details are described in \cite{ciotti17}. Briefly, the radiation pressure force per unit mass is given by
\begin{align}
    \left(\nabla p_\mathrm{rad}\right)_\mathrm{photo} & = -\frac{H}{c}\bm{e}_r,                                                     \\
    \left(\nabla p_\mathrm{rad}\right)_\mathrm{es}    & = -\frac{\rho \kappa_\mathrm{es}}{c}\frac{L_\mathrm{BH}}{4\pi r^2}\bm{e}_r,
\end{align}
where $\rho$ is the gas density and $\kappa_\mathrm{es}=0.35\,\mathrm{cm^2\,g^{-1}}$ is the electron-scattering opacity.

\subsubsection{Star formation and stellar feedback}\label{sec:star}
The star formation model follows that of \cite{yuan18}, but we restricted star formation to gas with densities exceeding $1\,\mathrm{cm^{-3}}$ and temperatures below $4 \times 10^4\,\mathrm{K}$ \citep{zhu23a}. The stellar feedback model is analogous to that of \cite{ciotti12}, incorporating mass loss from stellar winds as well as feedback from Type Ia and Type II supernovae. However, the rate of Type Ia supernovae in the \texttt{M32 model} was reduced to $0.13$ per century per $10^{10}\,L_\odot$ in the B band \citep{cappellaro97, luis03}. In addition, Type Ia supernovae were stochastically injected following a Poisson process, with the expected number set by the local explosion rate multiplied by the timestep \citep{zhu23b}.

\subsubsection{Galaxy models and setup}\label{sec:galaxy}
The galaxy model in MACER consists of three primary components: gas, stars, and a central supermassive black hole. The stellar component transfers mass and energy to the gas through stellar feedback processes, though the stars are stationary. Conversely, gas is consumed to form new stars, removing mass, momentum, and energy from the gaseous phase. The black hole, located at the galactic center, grows in mass through accretion from the surrounding gas and provides feedback to the surrounding medium through AGN radiation and winds. The dark matter halo is included only as an external gravitational potential and its evolution is not considered.

For both models, the initial gas distribution is assumed to be a hot, low-density, single-phase, fully-ionized medium in hydrostatic equilibrium within the specific gravitational potential of each galaxy model. The gas number density profile $n(r)$ is characterized as the beta model \citep{mo10}
\begin{align}
    n(r) = n_0 \left(1 + \frac{r^2}{r_c^2} \right)^{-1.5\beta},
\end{align}
where $n_0 = 0.25\,\mathrm{cm^{-3}}$ is the central number density, $r_c$ is the scale radius, and the slope parameter $\beta = 2/3$ is based on the observation \cite{anderson13}.

In the \texttt{fiducial model}, following the approach of \cite{ciotti09}, the initial stellar distribution $\rho_\star(r)$ is described as the Jaffe profile \citep{jaffe83},
\begin{align}
    \rho_\star(r)=\frac{M_\star r_J}{4\pi r^2(r_J+r)^2},
\end{align}
where the initial stellar mass is $M_\star = 3 \times 10^{11}\,M_\odot$ and the Jaffe radius $r_J = r_c = 6.9\,\mathrm{kpc}$ in our simulation. The corresponding effective radius is $r_e = 9.04\,\mathrm{kpc}$, the stellar velocity dispersion is $\sigma_0 = 260\,\mathrm{km\,s^{-1}}$, and the black hole initial mass is $M_\text{BH} = 1.8 \times 10^9\,M_\odot$, chosen to satisfy the Faber-Jackson relations \citep{faber76} and the $M_\text{BH}-\sigma$ relation \citep{kormendy13}. The simulation radial regime ranges from $2.5\,\mathrm{pc}$ to $250\,\mathrm{kpc}$.
To optimize computational efficiency, we first ran a $4\,\mathrm{Gyr}$ simulation, saving snapshots every $12.5\,\mathrm{Myr}$. After the system reached a quasi-steady or quasiperiodic state (after about $2.5\,\mathrm{Gyr}$), we selected a representative snapshot just prior to a strong wind event as the restart point.

In the \texttt{M32 model}, we adopted the I-band light profile from \cite{graham09}, which included an inner nuclear component, a Sérsic bulge, and an outer exponential envelope \citep{graham02}. The corresponding stellar profile of M32 was obtained by combining these components with their stellar $M/L$ ratios \citep{graham09, cappellari06, howley13}. The stellar velocity dispersion and the rotational support were adopted from \cite{howley13}. The initial black hole mass was set to $M_\text{BH} = 2.5\times 10^6\,M_\odot$ \citep{marel98, verolme02, luis03}. The simulation radial domain extended from $0.01\,\mathrm{pc}$ to $1\,\mathrm{kpc}$ (corresponding to $258''$). The dark matter halo of M32 was not included, as it may have been stripped by M31. Moreover, within the inner region ($< r_e \approx 30''$, \cite{cappellari06}), the absence of a dark matter halo could not be ruled out \citep{howley13}. The scale radius of the gas number density profile was set to $r_c=0.1\,\mathrm{kpc}$. The simulation was run for $10\,\mathrm{Myr}$, with snapshots saved every $0.01\,\mathrm{Myr}$. Given this short simulation timescale, the stellar distribution and black hole mass did not evolve significantly.

The differences in key parameters of the two galaxy models are summarized in Table~\ref{tab:models}. Besides, the accretion mode, wind parameters are scale-free, and the ratio of radial adjacent grid spacing, the boundary conditions and star formation criteria are the same for both models.
\begin{table*}[htbp]
    \centering
    \caption{Differences in parameters of the two galaxy models used in this work.}
    \label{tab:models}
    \begin{tabular}{lcccccc}
        \hline
        Model             & $M_\star$ ($M_\odot$) & Stellar profile                                                                 & $M_\text{BH}$ ($M_\odot$) & $r_\text{in}$ (pc) & $r_\text{out}$ (kpc) & \begin{tabular}{c}SNIa Rate \\($10^{-12}\,L_\odot^{-1}yr^{-1}$)\end{tabular} \\
        \hline
        \texttt{fiducial} & $3\times10^{11}$      & Jaffe                                                                           & $1.8\times10^9$           & 2.5                & 250                  & 0.32                                                                                                          \\
        \texttt{M32}      & $4\times10^8$         & \begin{tabular}{c}nuclear + Sérsic bulge \\ + exponential envelope\end{tabular} & $2.5\times10^6$           & 0.01               & 1                    & 0.13                                                                                                          \\
        \hline
    \end{tabular}
\end{table*}

\subsection{Particle acceleration at the shocks}\label{sec:cr}
Our approach for calculating synchrotron emission from wind-driven shocks involves two main steps. First, we post-processed the hydrodynamic simulation data to identify shocks and determined their properties, including shock strength and the associated compression ratios. This requires adapting shock detection algorithms to work effectively in multidimensional simulations, where shock fronts can have complex geometries.

Once shock fronts were identified, we applied diffusive shock acceleration theory to compute the energy spectrum of nonthermal electrons accelerated at these shocks, following the approach of \cite{jiang10}. We assume that a small fraction of electrons crossing the shock front are injected into the acceleration process and subsequently develop power-law energy distributions. The resulting electron population then radiated via synchrotron emission as it evolved under the combined effects of radiative cooling and adiabatic expansion. We further assume that these nonthermal electrons remain frozen within the fluid elements and advect with the fluid flow, as the gas temperatures in post-shocked regions remain above $10^4\,\mathrm{K}$, ensuring a nearly fully ionized state. This framework enables us to predict both the spatial and spectral properties of the observable radio emission.

\subsubsection{Shock detection}\label{sec:shock}
Unlike the one-dimensional case in \cite{jiang10}, accurately identifying the shock front is more challenging in multidimensional simulations \citep{wu13}. We adapt the method of \cite{lovely1999} to detect the shock front. First, we approximate the normal direction of the shock front by computing the pressure gradient. We then calculate the pre- and post-shock velocities as $u_1 = \bm{v}_1\cdot\frac{\nabla P_1}{\left|\nabla P_1\right|}$ and $u_2 = \bm{v}_2\cdot\frac{\nabla P_2}{\left|\nabla P_2\right|}$, where $P$ refers to gas pressure, $\bm{v}$ is gas velocity in laboratory reference frame, and subscripts 1 and 2 refer to the upstream and downstream of the shock wave respectively.
The shock speed is then derived from the Rankine-Hugoniot jump conditions as
\begin{align}
    u_\text{sh} = u_1+a_1\sqrt{\frac{\gamma+1}{2\gamma}\frac{P_2}{P_1}+\frac{\gamma-1}{2\gamma}},
\end{align}
where $a_1$ is the sound speeds upstream and we adopt an adiabatic index $\gamma=5/3$, since the gas in the shock region is predominantly hot and fully ionized. For a valid shock, the pre- and post-shock velocities and the shock speed must satisfy the entropy condition,
\begin{align}
    u_2+ a_2 > u_\text{sh} > u_1+ a_1.
\end{align}

\begin{figure}[htbp]
    \centering
    \includegraphics[width=\columnwidth]{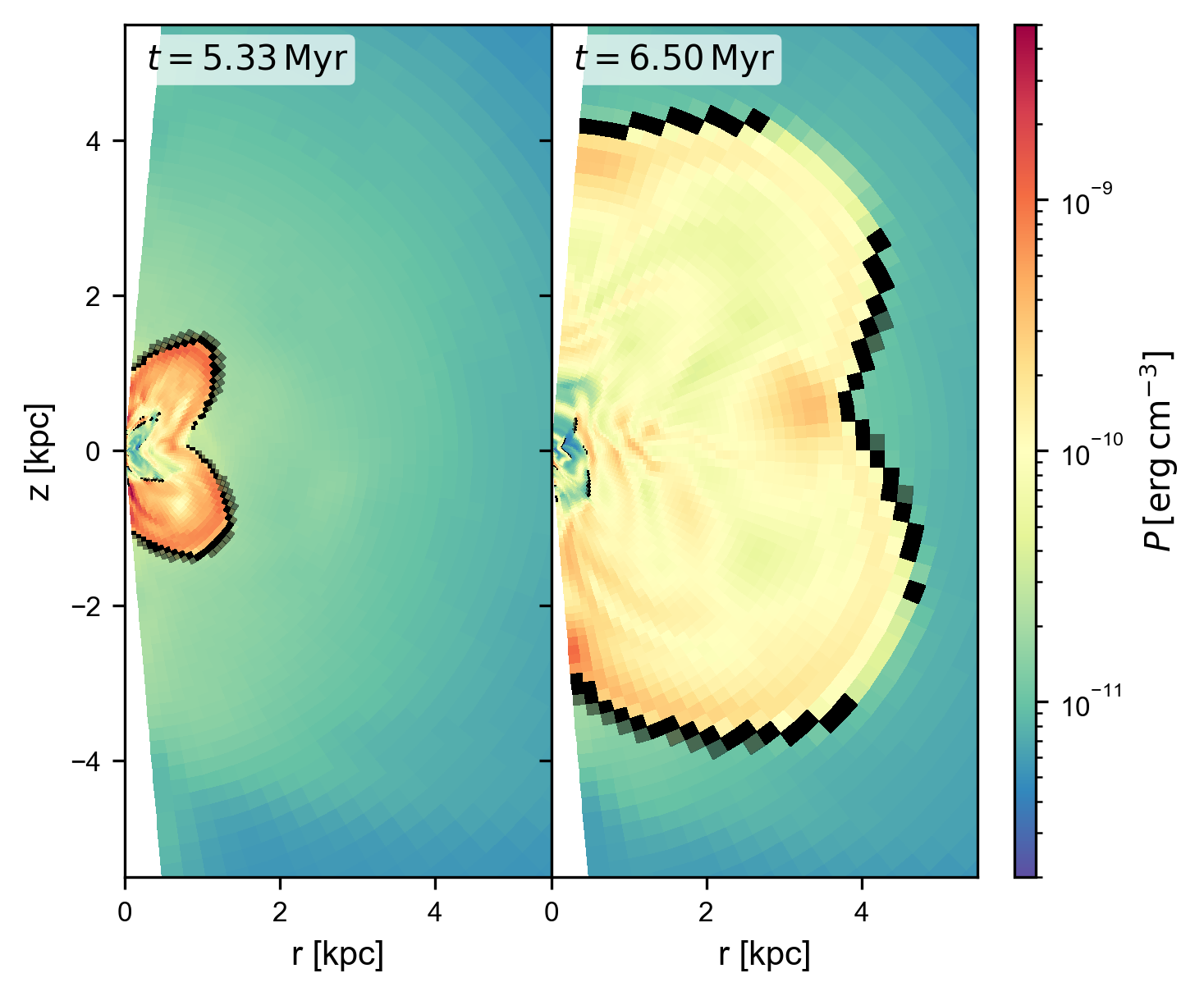}
    \caption{Shock detection in \texttt{fiducial model}. Color map shows pressure distribution at $t=5.33\,\mathrm{Myr}$ and $r=6.50\,\mathrm{kpc}$. Grayscale regions present detected shock fronts. Black regions mark post-shock zones with nonthermal electron injection.}
    \label{fig:shock_detect}
\end{figure}

To exclude very weak shocks, we impose a threshold condition $|\Delta P|/P > \varepsilon$. Fig.~\ref{fig:shock_detect} shows the detected shock fronts and post-shock zones where nonthermal electrons are injected. The value we use here, $\varepsilon = 2$, is for the purpose of demonstrating the process of a shock wave weakening from a strong shock. Note that this filter is primarily designed for low-resolution grids. Importantly, this criterion does not explicitly depend on the grid resolution. As discussed in \citet{lovely1999}, the choice of threshold is a relative concept; in our case, it mainly serves to filter out spurious signals in the outer regions of the galaxy, where large pressure differences across grid cells can arise from gravitational effects rather than genuine shocks, especially when the resolution is poor. To assess the impact of resolution on our shock detection algorithm, we have performed a Sod shock tube test \citep{sod78}, as described in the Appendix~\ref{app:shock_detection}. Although this filtering criterion is not strictly rigorous -- since Appendix~\ref{app:particle_spectrum} adopts more stringent thresholds for strong shocks -- it remains useful for algorithm validation and the identification of weak shocks. Note that, however, as the grid size becomes coarser, especially at larger radii in our simulations, shock waves may be smoothed out, whose intensities are thereby underestimated.

\subsubsection{nonthermal electron acceleration and synchrotron emission}\label{sec:spectrum}
Once shock fronts are identified, we model the acceleration of nonthermal electrons using the framework of diffusive shock acceleration theory \citep{Blandford87, reynolds08}. Our approach considers several key physical processes and assumptions:

\paragraph{Electron Injection and Acceleration:} We assume that a small fraction (approximately $10^{-3}$ to $10^{-4}$) of thermal electrons crossing the shock front are injected into the acceleration process \citep{drury89,berezhko94,kang95}. These electrons are accelerated via first-order Fermi acceleration, developing power-law momentum distributions with spectral indices determined by the shock compression ratio and upstream Mach number. Only strong shocks are considered, as weaker shocks cannot efficiently accelerate particles to relativistic energies. The detailed mathematical formulation of the electron injection efficiency and spectral index as functions of shock parameters is provided in Appendix~\ref{app:particle_spectrum}.

\paragraph{Energy Loss Mechanisms:} The accelerated nonthermal electrons lose energy through multiple channels: (1) synchrotron radiation and inverse Compton scattering off cosmic microwave background photons, and (2) adiabatic expansion as the shocked gas expands and cools \citep{reynolds98}. We assume these electrons remain frozen within fluid elements and advect with the gas flow, assuming that turbulent amplification processes are negligible, following the approach of \cite{reynolds98}. We refer readers to Appendix~\ref{app:synchro} for the detailed mathematical treatment of these energy loss processes.

\paragraph{Magnetic Field Model:} The magnetic field strength is parameterized through the plasma beta parameter $\beta \equiv P_\text{gas}/(B^2/8\pi) \approx 10$ \citep{churazov08, jiang10, wang20, wang21}, where $B$ is the magnetic field strength. This assumption allows us to relate the local magnetic field to the thermal gas pressure from our hydrodynamic simulations.

\paragraph{Maximum Electron Energy:} The maximum energy attained by accelerated electrons is determined by balancing the acceleration timescale against cooling losses. Electrons can be accelerated until synchrotron cooling becomes more rapid than the shock acceleration process. The maximum Lorentz factor $\gamma_\text{max}$ is estimated by equating the acceleration timescale to the synchrotron cooling timescale \citep{drury83,reynolds98, jiang10}, as shown in Eqs.~\ref{Gamma_max_app}.

\paragraph{Synchrotron Emission Calculation:} The resulting synchrotron emission is computed by integrating over the evolving nonthermal electron energy distribution, accounting for the temporal evolution of both the electron population and the local magnetic field. We track the spectral aging of the electron population as it moves away from the shock fronts and cools radiatively \citep{reynolds98, jiang10} as shown in Eqs.~\ref{Energy_spectrum_app} and~\ref{Synchrotron_emission_app}.

\section{Results}\label{sec:results}
Section~\ref{sec:result_fiducial} presents comprehensive results from \texttt{fiducial model}, including detailed analyses of the temporal evolution of synchrotron emission spectra, the spatial distribution of radio emission, and simulated radio telescopes observations. Section~\ref{sec:result_m32} focuses on analogous results for the specific case of \texttt{M32}, enabling a direct comparison with observational data.

\subsection{Synchrotron emission in a massive elliptical galaxy}\label{sec:result_fiducial}
\begin{figure*}[htbp]
    \centering
    \includegraphics[width=0.8\textwidth]{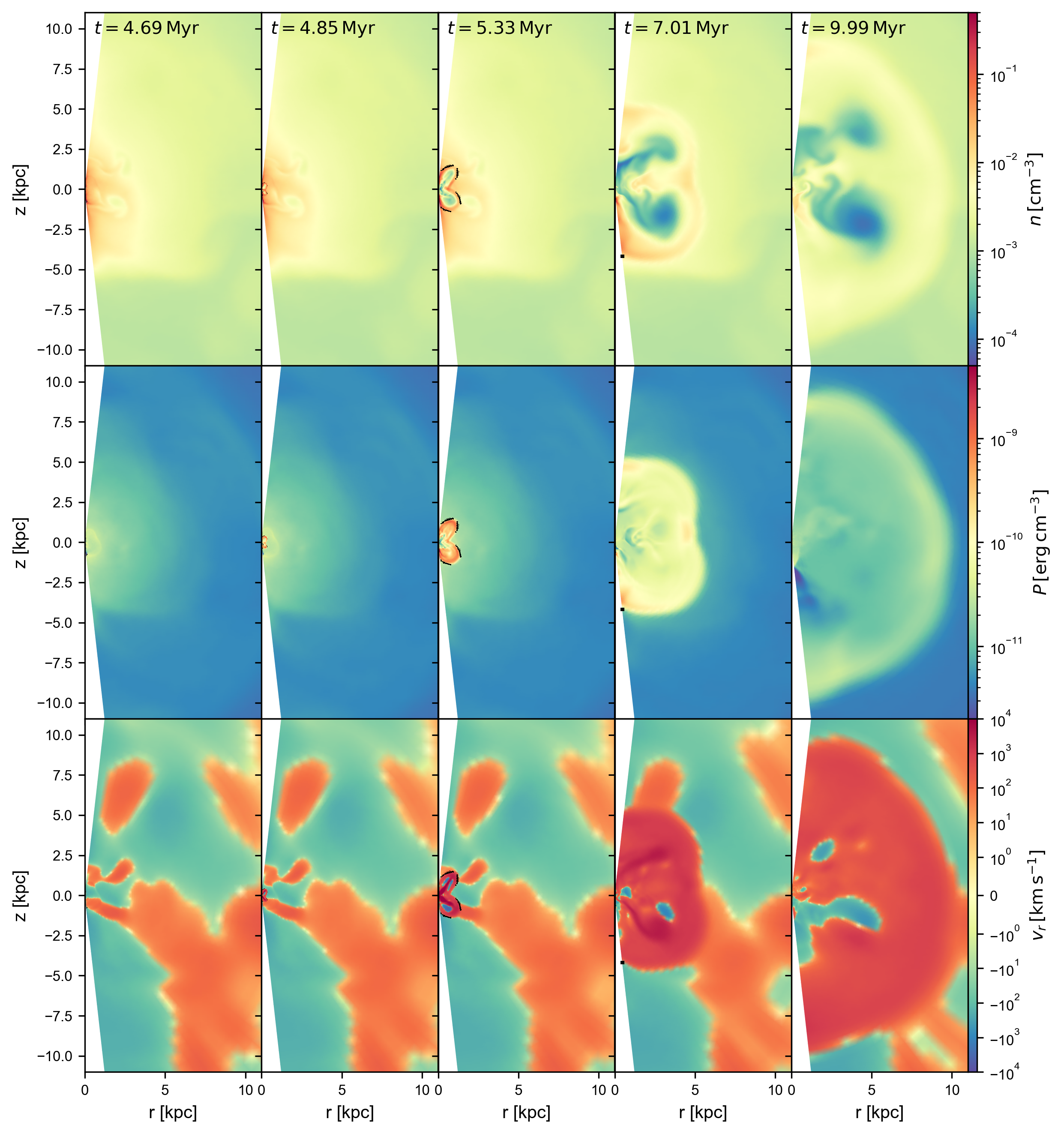}
    \caption{AGN winds and shocks evolution in \texttt{fiducial model}. Snapshots are shown at $t=4.69$, $4.75$, $5.50$, $7.01$, and $9.99\,\mathrm{Myr}$ (left to right). Top to bottom panels show density ($n$), pressure ($P$), and radial velocity ($v_r$) in the $r$-$z$ plane. Black regions mark strong shock fronts.}
    \label{fig:shock_evolution}
\end{figure*}

As outlined in Section~\ref{sec:hydro}, we conducted detailed numerical simulations to examine the temporal evolution of shock waves driven by AGN winds in a massive elliptical galaxy. The intensity of shock waves and radio emissions mainly depends on the power of the wind. The power of hot winds varies greatly depending on the accretion rate, while cold winds are generally stronger. What we show here is a shock wave caused by a strong hot wind. Fig.~\ref{fig:shock_evolution} shows the temporal evolution of wind-driven shocks. At $t=4.69\,\mathrm{Myr}$, near the onset of shock formation, the hot wind emerges from the central region with a bipolar geometry, characterized by high temperature and low density. As the wind expands, it initially drives strong shocks into the surrounding medium, producing a complex multiphase structure. By $t=5.33\,\mathrm{Myr}$, the wind-driven bubble expands to several kiloparsecs, though the shock waves already begin to weaken significantly. In addition, Fig.~\ref{fig:shock_detect} indicates the presence of shock flags around the low-pressure zone in the inner region. It turns out that the cold wind was injected from $t = 4.97\,\mathrm{Myr}$ to $t = 5.5\,\mathrm{Myr}$, which contributes to the inner shock. However, the inner shock never catches up with the outer major shock before dissipating, so we believe that it does not amplify the major shock. The later stages ($t>7.01\,\mathrm{Myr}$), instabilities develop along the wind-ISM interface, and the shock fronts weaken further, evolving into weak waves. By $t=9.99\,\mathrm{Myr}$, the outflow develops into a large-scale structure extending to about $10\,\mathrm{kpc}$, exhibiting complex density and velocity patterns that indicate significant mixing between the wind and ambient medium, while the shocks largely dissipate into sound waves. Note that in this event, the presence of a cold wind may result in the displacement of hot wind material to a greater distance. In our model, nonthermal electrons are injected at post-shock positions.

\begin{figure}[htbp]
    \centering
    \includegraphics[width=\columnwidth]{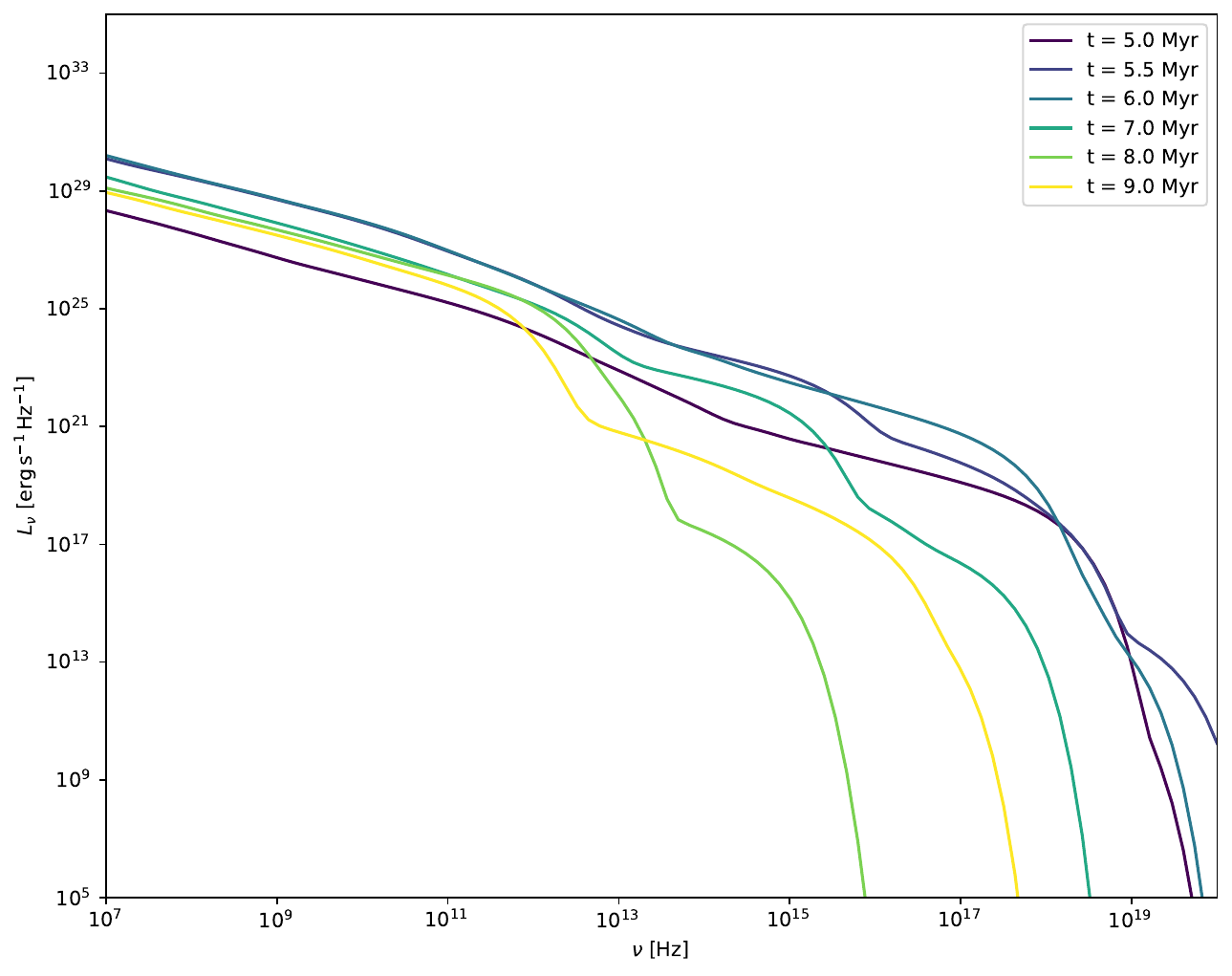}
    \caption{Synchrotron emission spectra in \texttt{fiducial model}. Colored lines show different times from $t=5.0\,\mathrm{Myr}$ (purple) to $t=9.0\,\mathrm{Myr}$ (yellow). Spectra show broken power-law shapes with high-frequency cutoffs.}
    \label{fig:spectrum_evolution}
\end{figure}

As shown in Fig.~\ref{fig:spectrum_evolution}, the emission spectrum exhibits a broken power-law shape with a steep cutoff at high frequencies. With cold wind injected from $t=4.9\,\mathrm{Myr}$ to $t=5.5\,\mathrm{Myr}$, inner and outer shock waves propagate and generate significant numbers of nonthermal electrons, leading to enhanced synchrotron luminosity. Subsequently, the luminosity gradually declines while the cutoff frequency decreases, reflecting both the cooling and dissipation of shock-accelerated electrons and the continued weakening of the shock waves.

\begin{figure}[htbp]
    \centering
    \includegraphics[width=\columnwidth]{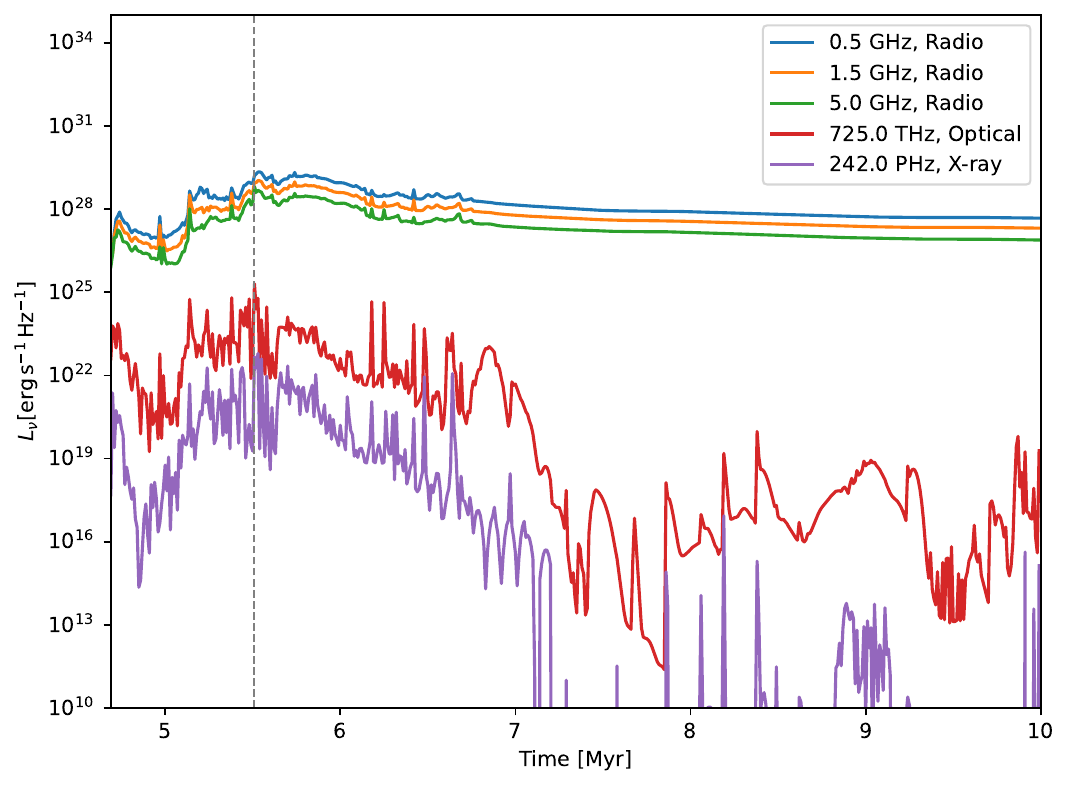}
    \caption{Synchrotron light curves in \texttt{fiducial model}. The blue, orange and green lines present radio emissions in $0.5$, $1.5$ and $5.0\,\mathrm{GHz}$ respectively. The red line show the optical emission in $725.0\,\mathrm{THz}$. The purple line show the X-ray emission in $242.0\,\mathrm{PHz}$. Vertical dashed line at $t=5.5\,\mathrm{Myr}$ marks shock weakening onset.}
    \label{fig:lightcurve}
\end{figure}

To illustrate the temporal evolution of the emission across different wavebands, we track the luminosity at several representative frequencies (Fig.~\ref{fig:lightcurve}). The radio emission at $0.5$, $1.5$, and $5.0\,\mathrm{GHz}$ shows relatively smooth evolution, characterized by a gradual rise as the shock waves expand, peaking at luminosities of $\sim 10^{29}\,\mathrm{erg\,s^{-1}\,Hz^{-1}}$ around $t=5.5\,\mathrm{Myr}$, followed by a slow decline. In contrast, the optical and X-ray emission display more pronounced variability, with luminosities fluctuating by several orders of magnitude, indicating their sensitivity to the highest-energy electrons. The vertical dashed line at $t=5.5\,\mathrm{Myr}$ marks a critical transition, beyond which the major wind-driven shock wave begins to weaken significantly. Notably, even after $7\,\mathrm{Myr}$, significant X-ray emission persists due to ongoing AGN activity as the gas cools, accretion continues, driving new shocks near the galactic center that contribute to sustained high-energy emission. The total luminosity thus reflects contributions from both the major wind event and subsequent smaller shock waves.

To quantitatively characterize the synchrotron spectrum, we calculate the power-law index $\alpha$ (where $L_\nu \propto \nu^\alpha$) in the radio band. The spectral index $\alpha$ is calculated as
\begin{align}
    \alpha = \frac{\log(L_{\nu_2}) - \log(L_{\nu_1})}{\log(\nu_2) - \log(\nu_1)},
\end{align}
where the frequencies $\nu_1$ and $\nu_2$ are taken to be $0.5$ and $5.0\,\mathrm{GHz}$, respectively.

\begin{figure*}[htbp]
    \centering
    \includegraphics[width=0.8\textwidth]{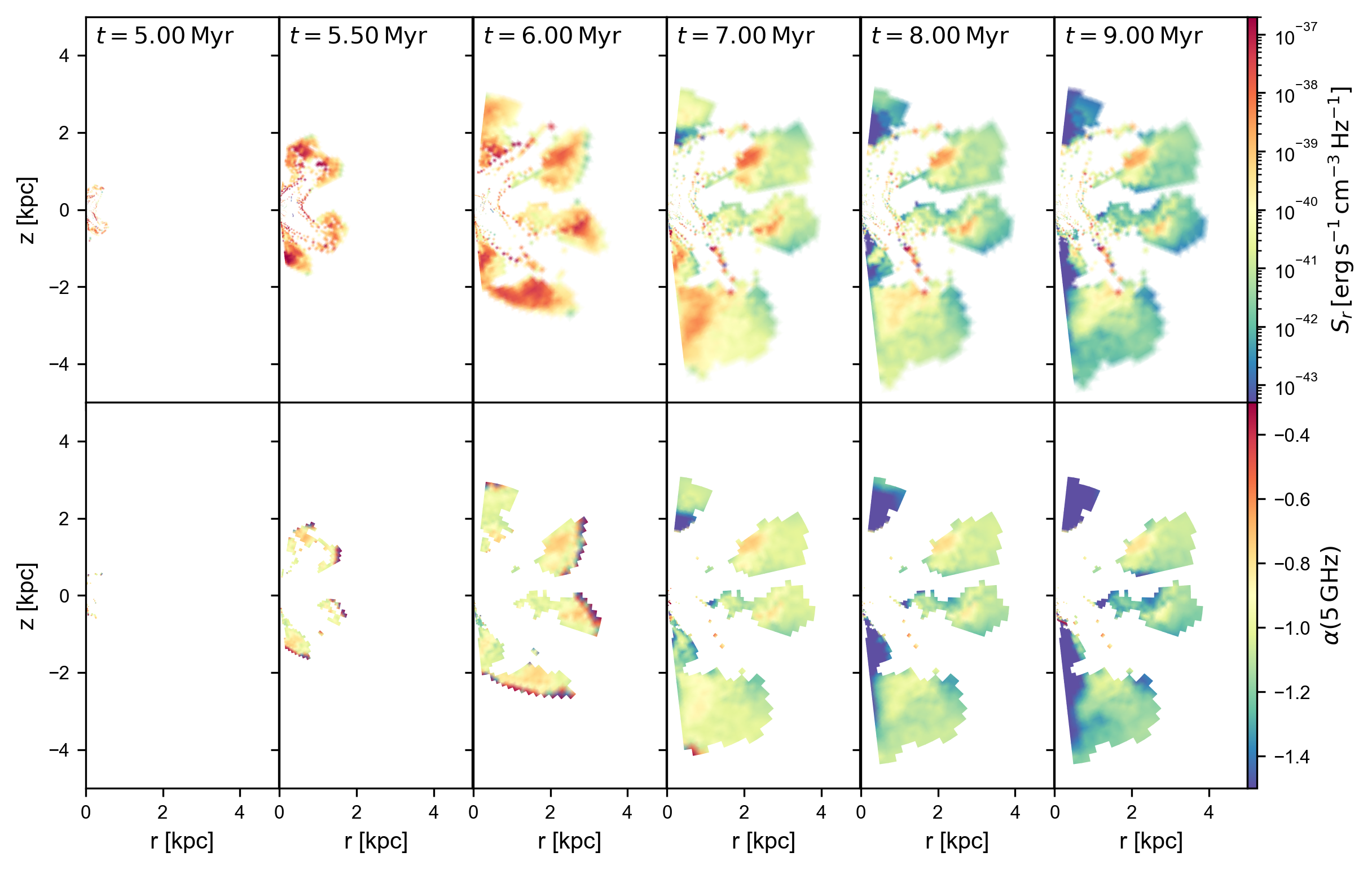}
    \caption{Spatial distribution of synchrotron emission properties in \texttt{fiducial model}. Top panels show $5.0\,\mathrm{GHz}$ emissivity ($\mathrm{erg\,s^{-1}\,cm^{-3}\,Hz^{-1}}$). Bottom panels show spectral index $\alpha$ ($L_\nu \propto \nu^\alpha$ between $0.5-5.0\,\mathrm{GHz}$). Times are $t=5.0$, $5.5$, $6.0$, $7.0$, $8.0$, $9.0\,\mathrm{Myr}$ (left to right).}
    \label{fig:alpha_spatial}
\end{figure*}

\begin{figure}[htbp]
    \centering
    \includegraphics[width=\columnwidth]{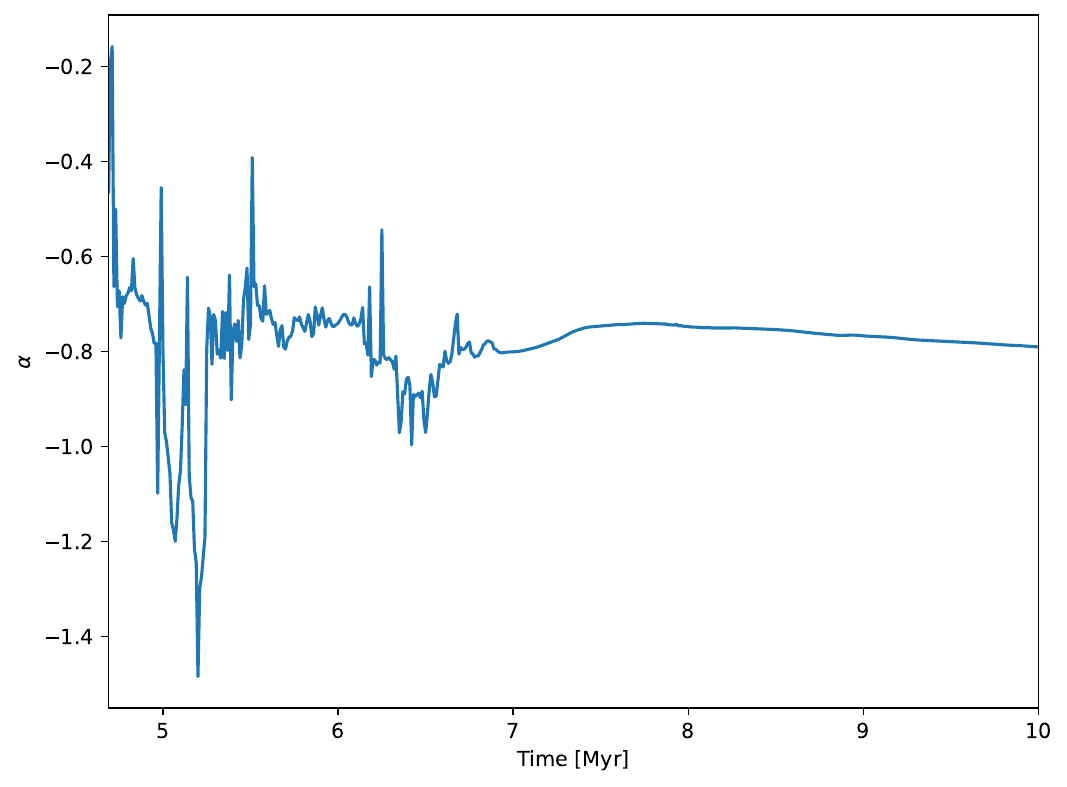}
    \caption{Volume-integrated radio spectral index evolution in \texttt{fiducial model}. Initial fluctuations occur at $t < 5.5\,\mathrm{Myr}$ followed by stabilization around $\alpha \approx -0.8$.}
    \label{fig:alpha}
\end{figure}

Fig.~\ref{fig:alpha_spatial} presents the spatial distributions of the synchrotron emissivity at $5.0\,\mathrm{GHz}$ (top panels) and the spectral index $\alpha$ (bottom panels) at several stages of the AGN-driven outflow evolution. At early times ($t=5.0\,\mathrm{Myr}$), the emission is concentrated in small, discrete regions near the strongest shock fronts, characterized by high emissivities and relatively flat spectral indices ($\alpha \approx -0.4$), indicating efficient particle acceleration at strong shocks and a correspondingly hard electron energy spectrum. As the outflow expands ($t=5.5$-$6.0\,\mathrm{Myr}$), the emitting regions grow more extended. Interestingly, the flattest spectral indices remain near the shock fronts, where particle acceleration is most efficient, although the number density of accelerated electrons may be decreasing. This spatial offset between peak emissivity and the flattest spectral index reflects the complex interplay between shock strength, particle acceleration efficiency, and the evolving population of nonthermal electrons. At later times ($t>7.0\,\mathrm{Myr}$), the primary shock weakens significantly (Fig.~\ref{fig:shock_evolution}), and the emission spreads over a larger volume, with reduced intensity and steeper spectral indices ($\alpha \approx -1.0$ to $-1.4$), indicating both aging of the electron population and weakening of the shocks. However, in the innermost regions ($r<0.5\,\mathrm{kpc}$), the spectral index remains relatively flat ($\alpha \approx -0.4$), suggesting ongoing particle acceleration by shocks associated with newly emerging minor winds. This behavior is consistent with previous observations \citep{kaiser07}. The temporal evolution of the volume-integrated spectral index (Fig.~\ref{fig:alpha}) reveals the competition between the gain and loss of nonthermal electron total energy. During the early phase of shock evolution ($t < 5.5\,\mathrm{Myr}$), the spectral index varies significantly, with values ranging from $-1.5$ to $-0.2$. These fluctuations reflect the dynamic interplay between particle acceleration and cooling processes as the shock waves develop and intensify. After $t = 7\,\mathrm{Myr}$, the spectral index gradually stabilizes around $-0.8$. This steeper spectrum is consistent with synchrotron aging of the electron population in the post-shock regions, where radiative losses outweigh continued particle acceleration.

\begin{figure}[htbp]
    \centering
    \includegraphics[width=\columnwidth]{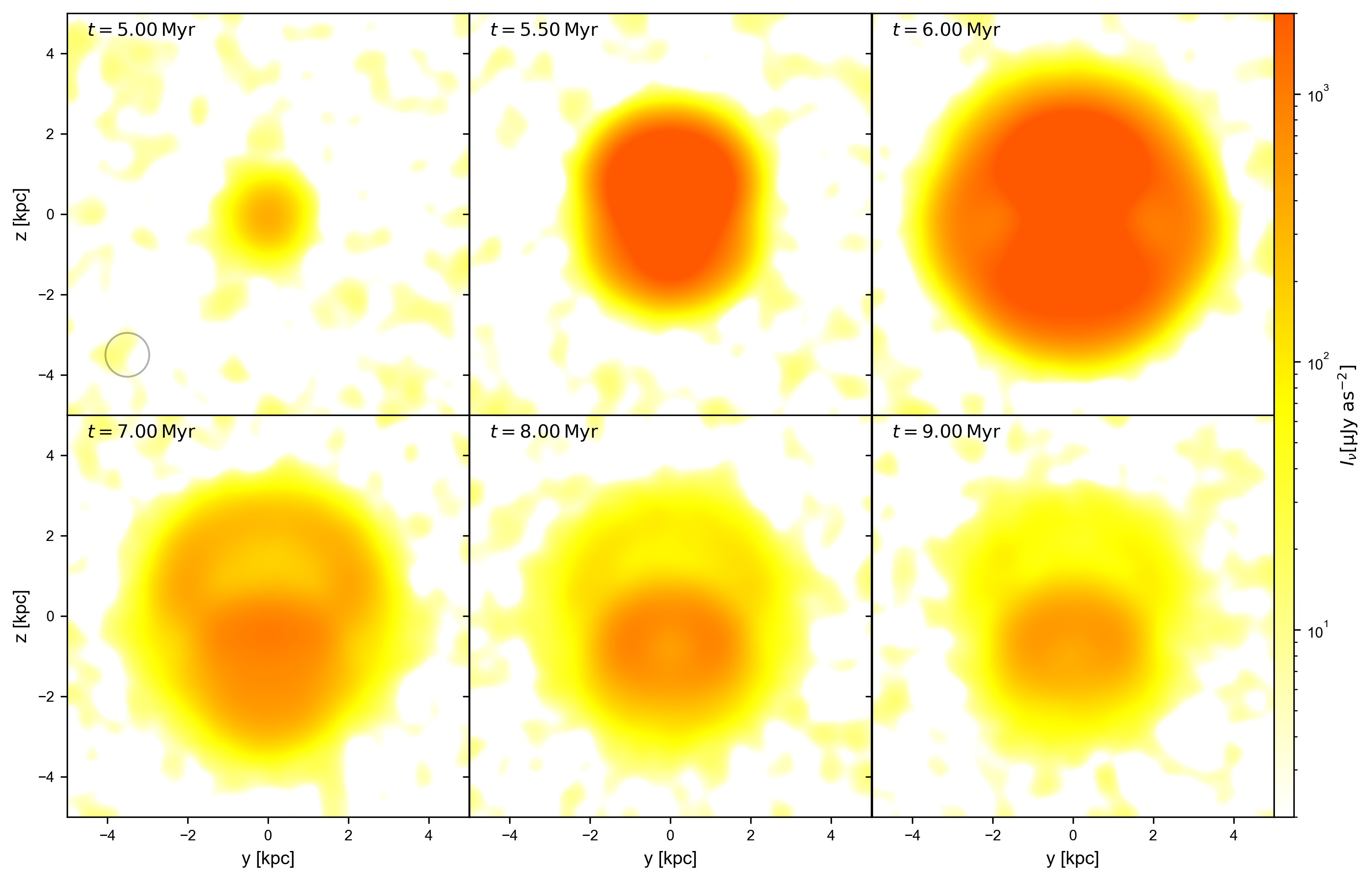}
    \caption{Simulated $5\,\mathrm{GHz}$ radio observations of \texttt{fiducial model}. Surface brightness is shown in $\mathrm{\mu Jy\,as^{-2}}$ viewed at $45^\circ$ inclination for a source at $50\,\mathrm{Mpc}$. Beam size is $4.5''$ (gray circle) with noise level $4.65\,\mathrm{\mu Jy/beam}$.}
    \label{fig:radio_observation}
\end{figure}

To predict the appearance of these synchrotron structures in future radio observations, we simulate their observation with next-generation radio telescopes. Assuming axial symmetry in our 2D simulation data, we construct a 3D model by rotating the data around the z-axis and projecting the emission along a line of sight with a pitch angle of $45^\circ$. Fig.~\ref{fig:radio_observation} presents the simulated $5\,\mathrm{GHz}$ radio images at various evolutionary stages, assuming the source is located at a distance of $50\,\mathrm{Mpc}$. The images are convolved with a $4.5''$ beam and include a realistic noise level of $4.65\,\mathrm{\mu Jy/beam}$, consistent with the sensitivity achievable in a one-hour SKA-Mid observation.

The simulated observations show that the radio emission evolves from a compact, moderately faint source at $t=5.0\,\mathrm{Myr}$ to an extended, diffuse structure by $t=9.0\,\mathrm{Myr}$. At $t=5.5\,\mathrm{Myr}$, when shock activity peaks, the radio source displays a bright core with a surface brightness exceeding around $10^4\,\mathrm{\mu Jy\,as^{-2}}$. As the outflow expands, the emission becomes more extended but diminishes in surface brightness. After $t=7.0\,\mathrm{Myr}$, emission from the outermost regions becomes undetectable due to the weakening of the primary shock, while the core remains visible owing to ongoing minor shocks in the central region.

\subsection{Synchrotron emission in M32}\label{sec:result_m32}
In addition to \texttt{fiducial model} of a massive elliptical galaxy, we apply our methodology to the nearby dwarf elliptical galaxy M32, as described in Section~\ref{sec:galaxy}. This allows a direct comparison between our theoretical predictions and observational constraints for a well-studied system. \texttt{M32 model} differs markedly from \texttt{fiducial model} in its physical properties, possessing a much smaller black hole mass ($M_{\rm BH} = 2.5 \times 10^6\,M_\odot$) and a shallower gravitational potential.

The hydrodynamic evolution of AGN-driven outflows in \texttt{M32 model} follows similar physical principles as in the \texttt{fiducial} case, but differs in scale and intensity. The lower black hole mass and accretion rate result in substantially reduced energy input from AGN winds, producing weaker shocks that dissipate more quickly. To facilitate comparison with observations and minimize contamination from supernova-driven shocks -- whose strength can rival wind-driven shocks at larger radii -- we restrict our analysis to the central $6'' \times 6''$ region, following \citet{peng20}. This restricted field of view helps disentangle AGN-driven and supernova-driven effects, as the latter typically expand to larger radii (tens of parsecs) before attaining similar strengths.

Given M32's currently low AGN luminosity \citep{luis03,yang15}, we select a representative event featuring a modest-luminosity hot wind that could generate moderately strong shocks in the past, after the system reaches a quasi-steady or quasiperiodic state ($\sim 1\,\mathrm{Myr}$). This approach enables exploration of a scenario in which M32 underwent a moderate AGN outburst sufficient to drive detectable shocks into the surrounding medium. Focusing on this episode enables more direct comparison with current observational constraints.

\begin{figure*}[htbp]
    \centering
    \includegraphics[width=0.8\textwidth]{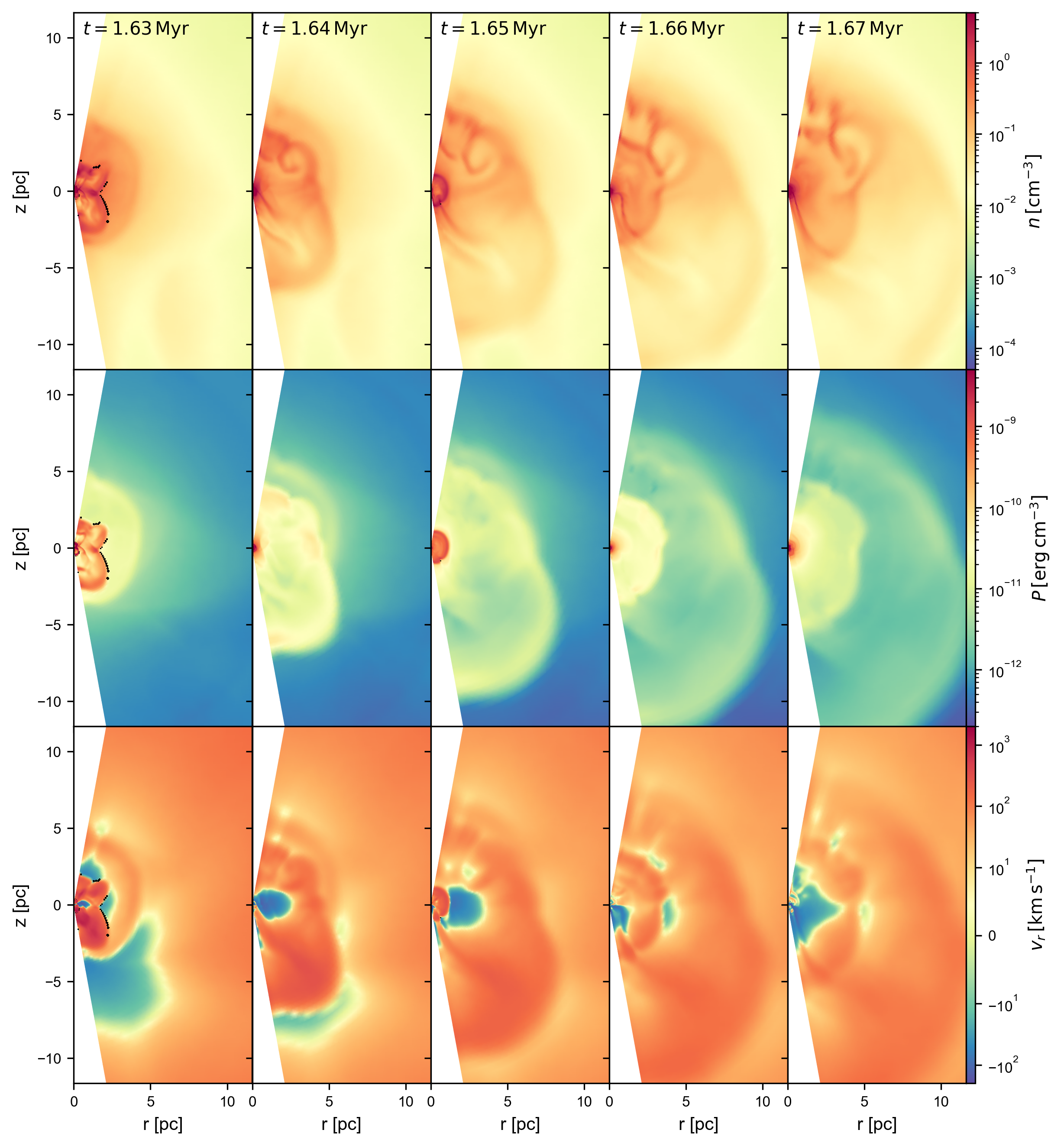}
    \caption{Same as Fig.~\ref{fig:shock_evolution} but for \texttt{M32 model}. Snapshots are shown at $t=1.63$, $1.64$, $1.65$, $1.66$, and $1.67\,\mathrm{Myr}$.}
    \label{fig:m32_shock_evolution}
\end{figure*}

The hydrodynamic evolution of a hot wind in \texttt{M32 model} during a representative outburst is shown in Fig.~\ref{fig:m32_shock_evolution}. The simulation spans $t = 1.63$ to $1.67\,\mathrm{Myr}$, capturing the development and propagation of shock waves through the central region. During this period, the AGN bolometric luminosity fluctuates between $10^{-9}$ and $10^{-6}\,L_\mathrm{Edd}$, broadly consistent with observational constraints.

In contrast to \texttt{fiducial model}, the shock waves in \texttt{M32 model} are substantially weaker and more spatially confined, typically extending only to about $10\,\mathrm{pc}$ from the center. This restricted extent results from the reduced energy input from M32's less massive central black hole. Moreover, the shocks in \texttt{M32} dissipate much more rapidly, with significant weakening occurring within approximately $0.04\,\mathrm{Myr}$, as opposed to several $\mathrm{Myr}$ in the \texttt{fiducial} case. Despite these differences in scale and intensity, the underlying physical mechanisms of shock formation and particle acceleration remain consistent, enabling us to apply our synchrotron emission model to predict the resulting radio emission.

\begin{figure}[htbp]
    \centering
    \includegraphics[width=\columnwidth]{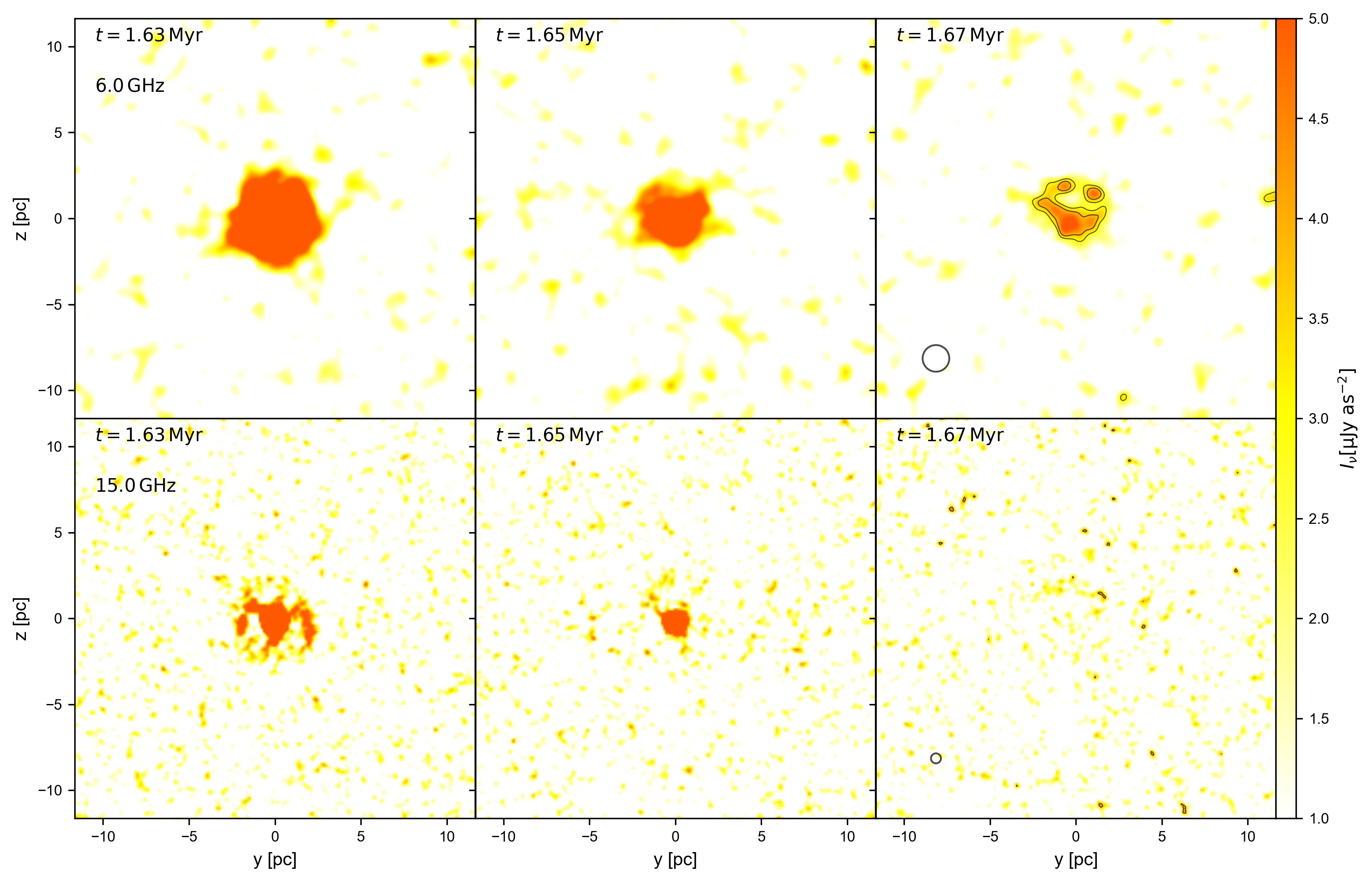}
    \caption{Simulated radio observations of \texttt{M32 model} inner $6''\times 6''$ region. Surface brightness is shown in $\mathrm{\mu Jy\,as^{-2}}$, viewed at $45^\circ$ inclination for M32 at $0.8\,\mathrm{Mpc}$. Top panels show $6.0\,\mathrm{GHz}$ with beam size $0.4''$ and noise level $1.0\,\mathrm{\mu Jy/beam}$. Bottom panels show $15.0\,\mathrm{GHz}$ with beam size $0.15''$ and noise level $1.3\,\mathrm{\mu Jy/beam}$. Times are $t=1.63$, $1.65$, and $1.67\,\mathrm{Myr}$ (left to right). Black contours in the right panels present $(3, 3.8)\times \mathrm{rms}$.}
    \label{fig:m32_radio_simulation}
\end{figure}

We simulated radio images of \texttt{M32} at various evolutionary stages of the AGN hot wind (Fig.~\ref{fig:m32_radio_simulation}). The top panels depict the predicted emission at $6.0\,\mathrm{GHz}$, while the bottom panels display the emission at $15.0\,\mathrm{GHz}$. The simulations assume M32's actual distance of $0.8\,\mathrm{Mpc}$ and incorporate realistic observational parameters, including beam sizes of $0.4''$ at $6.0\,\mathrm{GHz}$ and $0.15''$ at $15.0\,\mathrm{GHz}$ (indicated in the right panels), as well as noise levels of $1.0\,\mathrm{\mu Jy/beam}$ and $1.3\,\mathrm{\mu Jy/beam}$ respectively, consistent with the observations reported in \cite{peng20}.

At early times ($t=1.63\,\mathrm{Myr}$), the simulated radio emission in \texttt{M32} exhibits a compact, centrally concentrated morphology, with a peak surface brightness of approximately $10^4\,\mathrm{\mu Jy\,as^{-2}}$ at $6.0\,\mathrm{GHz}$. As the outflow evolves ($t=1.65\,\mathrm{Myr}$), the emission develops asymmetric structures, reflecting the complex shock patterns revealed in the hydrodynamic simulations. By the later stage ($t=1.67\,\mathrm{Myr}$), the emission weakens significantly and fragments into discrete regions, particularly at $15.0\,\mathrm{GHz}$, where the higher frequency increases sensitivity to the cooling of the highest-energy electrons. The radio morphology and intensity at $t=1.67\,\mathrm{Myr}$ are remarkably consistent with the R1 source observed by \cite{peng20}, suggesting that hot-wind-driven shocks could provide a plausible physical explanation for this radio component. However, we note that this comparison represents a specific evolutionary phase selected from our simulation, and such emission characteristics may not be persistent throughout the entire wind evolution cycle. To quantify this, we analyzed the simulation from the time the system reaches a quasi-steady state to the end of the run. We find that the system remains in a hot-mode state for approximately 89\% of the time. However, the fraction of time when the central region is both in a low-accretion state and not affected by new wind-driven shocks, i.e., when the simulated morphology matches the observed compact radio source, is only about 5\%. This suggests that the observed radio morphology is a transient feature, likely associated with the aftermath of a modest AGN wind event, and not a persistent state.

\section{Discussions}\label{sec:discussion}
Our simulations of synchrotron emission from wind-driven shocks in elliptical galaxies offer valuable insights into the radio signatures of AGN feedback and their observational consequences. In this section, we discuss the significance of our findings, compare them with observations, and outline prospects for future research.

\subsection{Comparison with observations and model validation}
The comparison of our simulated radio emission in \texttt{M32} with existing observational data serves as a crucial test of our theoretical framework. The morphology, intensity, and spectral properties of the simulated emission at $t=1.67\,\mathrm{Myr}$ exhibit remarkable consistency with the R1 radio source observed by \cite{peng20}. This agreement is particularly noteworthy, as our model makes specific predictions regarding the physical origin of this emission, specifically, that it originates from nonthermal electrons accelerated at shocks driven by a modest AGN hot wind event, without the need for high AGN luminosities. This interpretation aligns with the current low-luminosity state of M32's AGN, as we are observing the aftermath of a past wind event, rather than ongoing, powerful activity. However, note that our match represents a specific snapshot in the evolutionary sequence of wind-driven shocks, selected from a range of possible configurations during the simulation. Thus the agreement we observe could be coincidental.

For alternative origins, we note that radio emission from jets is typically characterized by a flat spectrum and is optically thick, resulting in generally low polarization degrees. In contrast, wind-driven synchrotron emission evolves toward a steep spectrum and is optically thin, with higher polarization expected, but actually the observed polarization may be significantly lower than one might expect due to the tangled magnetic field. Thus, polarization and spectral-index mapping can help distinguish between the two. For supernova remnant shocks, the early-time spectrum is also relatively flat, but old remnants usually have sizes of tens of parsecs, much larger than the few-parsec scale of the observed central source. X-ray binaries can produce compact, steady jets on smaller scales than \texttt{M32}, but transient jet outbursts may be difficult to distinguish from AGN-wind-related events. Therefore, high-resolution, multi-frequency, and polarization-sensitive observations are essential to discriminate between these scenarios.

For massive elliptical galaxies, direct observational validation for massive elliptical galaxies is more challenging due to the limited sensitivity of current radio facilities. Our predictions for spectral index evolution, from flat spectra ($\alpha \approx -0.4$ to $-0.6$) in regions of active particle acceleration to steeper spectra ($\alpha \approx -0.8$ to $-1.4$) in aging plasma, are consistent with observations of radio sources in elliptical galaxies \citep{kaiser07}. This agreement further strengthens confidence in the validity of our physical model.

\subsection{Observational prospects with next-generation radio telescopes}
Our simulations show that wind-driven shocks can produce detectable radio emission across a range of galaxy masses, from dwarf ellipticals such as M32 to massive ellipticals. However, detecting this emission, especially in its extended and diffuse form, requires the sensitivity and angular resolution capabilities of next-generation radio facilities.

\subsubsection{SKA observations}
The SKA will revolutionize our ability to detect and characterize faint, diffuse radio emissions in galaxies. Based on our simulations, we predict that SKA-Mid, operating at frequencies between $0.35$ and $15.4\,\mathrm{GHz}$, will be able to detect extended synchrotron emission from wind-driven shocks in massive elliptical galaxies up to distances of roughly $50-100\,\mathrm{Mpc}$ with reasonable integration times ($\sim 10\,\mathrm{hours}$). The exceptional sensitivity of SKA (with noise levels of $4.65\,\mathrm{\mu Jy/beam}$ at $5\,\mathrm{GHz}$ for a one-hour observation\footnote{calculated by \url{https://sensitivity-calculator.skao.int/mid}}) will allow for the detection of outer, low-surface-brightness emission regions that are currently inaccessible to existing facilities.

For nearby massive elliptical galaxies, SKA will provide unprecedented detail, potentially resolving the fine structure of shock fronts and enabling spatially resolved spectral index mapping. This capability will be crucial for distinguishing between various acceleration mechanisms, tracking the evolution of the electron population, and providing information of AGN feedback.

\subsubsection{FAST Core Array and other facilities}
The FAST Core Array consists of 24 40-meter antennas distributed within a five-kilometer radius of the FAST telescope. By 2027, the full array is expected to become operational, offering exceptional sensitivity for detecting faint radio emissions. With the full 24-antenna configuration, the FAST Core Array will achieve an imaging sensitivity of $0.91\,\mathrm{\mu Jy/beam}$ in just one hour at $1.4\,\mathrm{GHz}$, making it especially effective for detecting diffuse synchrotron emission from wind-driven shocks in massive elliptical galaxies out to distances of $50-100\,\mathrm{Mpc}$ \citep{jiang24}. The angular resolution at $1.4\,\mathrm{GHz}$ will be roughly $5''$, comparable to that of the VLA FIRST survey, enabling detailed mapping of radio structures in nearby elliptical galaxies. This combination of high sensitivity and moderate resolution makes the FAST Core Array an ideal instrument for studying the extended, low-surface-brightness emission characteristic of aging, shock-accelerated electron populations in elliptical galaxies.

Other facilities, such as the ngVLA and the upgraded Giant Metrewave Radio Telescope (uGMRT), will also play key roles in this field. The ngVLA's high angular resolution and sensitivity across frequencies of $1.2-116\,\mathrm{GHz}$ will be particularly valuable for detailed studies of the central regions of nearby galaxies \citep{ngVLA}, while the uGMRT's capabilities at lower frequencies ($50-1500\,\mathrm{MHz}$) will aid in constraining the spectral properties of older electron populations \citep{uGMRT}.

\subsubsection{Observational strategies}
Based on our simulations, we propose the following observational strategies for detecting and characterizing synchrotron emission from wind-driven shocks:

1. Multifrequency observations covering at least $0.5-15\,\mathrm{GHz}$ to constrain the spectral index and its spatial variations.

2. High angular resolution ($<5''$ for massive elliptical galaxies within $50\,\mathrm{Mpc}$) to resolve the spatial structure of the emission and differentiate between distinct components.

3. Deep integrations to achieve noise levels of about $5\,\mathrm{\mu Jy/beam}$ or better, necessary for detecting the diffuse emission in the outer regions.

\subsection{Model limitations and future improvements}
\subsubsection{Magnetic field assumptions}
\label{sec:magnetic_field_discussion}
The assumption of a constant plasma beta parameter \citep[$\beta \approx 10$,][]{churazov08, jiang10, wang20, wang21} throughout the simulation domain represents a significant simplification. In reality, the magnetic field strength and topology likely vary both spatially and temporally, driven by processes such as turbulent amplification, compression at shock fronts, and dynamo action \citep{federrath16,ji16,donnert18,hu22}.

\begin{figure}[htbp]
    \centering
    \includegraphics[width=\columnwidth]{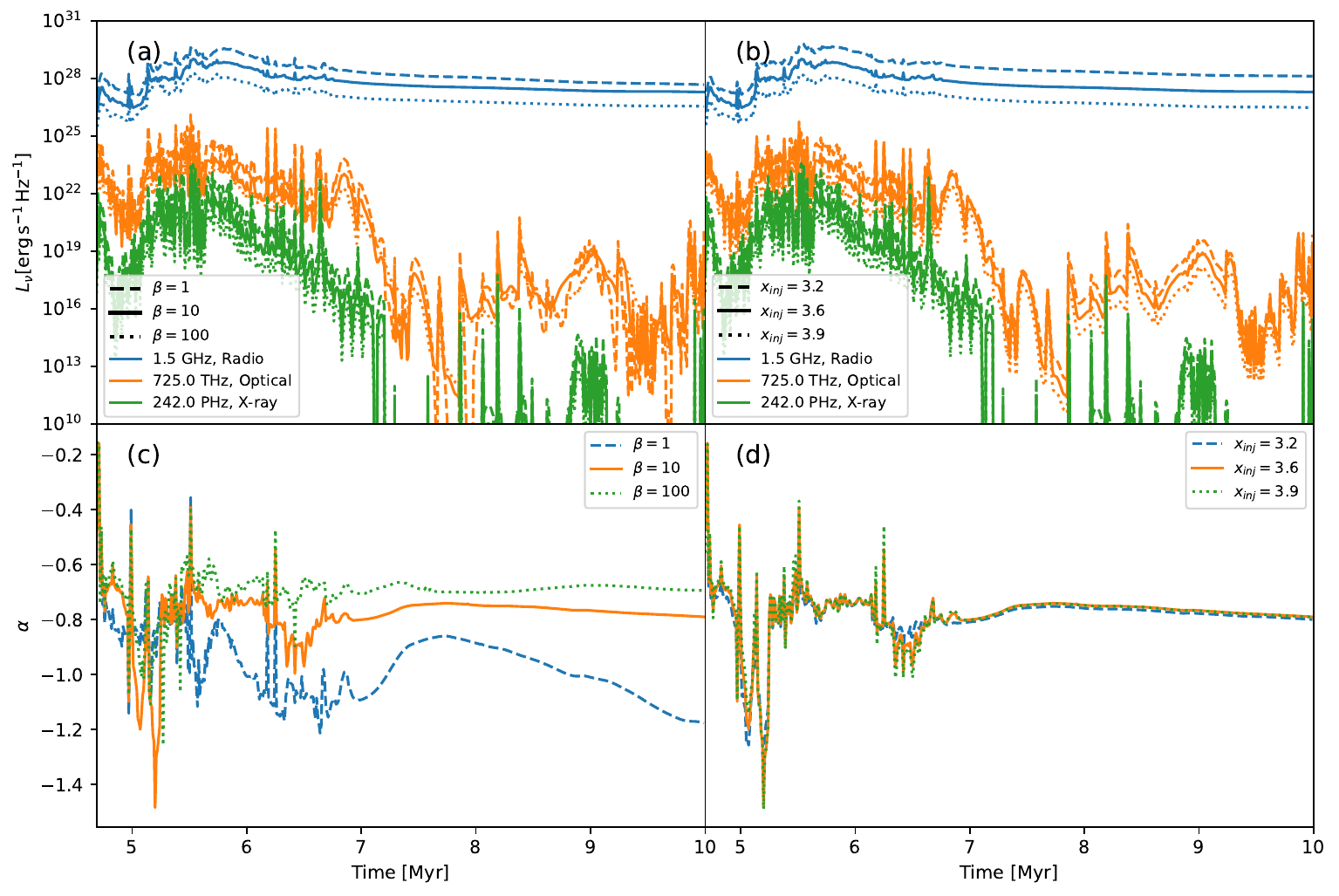}
    \caption{Synchrotron properties for different magnetic field strengths (left panels) and injection parameter $x_\text{inj}$ (right panels). Top panel shows luminosity evolution; the blue, orange, and green lines show radio, optical, and X-ray band respectively. Bottom panel shows spectral index evolution. The solid lines represent the fiducial model with $\beta = 10$ (left) or $x_\text{inj} = 3.6$ (right) , while the dashed and dotted lines represent $\beta = 1$ (left), $x_\text{inj} = 3.2$ (right) and $\beta = 100$ (left), $x_\text{inj} = 3.9$ (right) respectively}
    \label{fig:beta_comparison}
\end{figure}

To assess the impact of magnetic field strength on our results, we performed additional calculations with different plasma beta values ($\beta = 1$ and $\beta = 100$), corresponding to strong and weak magnetic field environments, respectively. Left panels in Fig.~\ref{fig:beta_comparison} compares the synchrotron luminosity evolution and spectral index evolution for different magnetic field strengths. For strong magnetic fields ($\beta = 1$), the synchrotron luminosity reaches higher peak values ($\sim 10^{30}\,\mathrm{erg\,s^{-1}\,Hz^{-1}}$) but evolves more rapidly, with faster cooling and faster spectral steepening. In contrast, weak magnetic fields ($\beta = 100$) produce lower peak luminosities ($\sim 10^{28}\,\mathrm{erg\,s^{-1}\,Hz^{-1}}$) but sustain the emission for longer periods, with more gradual spectral evolution maintaining relatively flat indices ($\alpha \sim -0.7$). At 1.5 GHz, the average luminosity for $\beta = 1$ is approximately 3.6 times higher than that of the fiducial model ($\beta = 10$), while the fiducial model's luminosity is about 5.5 times higher than that of $\beta = 100$. Similarly, the surface brightness comparisons follow the same trend.

These results demonstrate that magnetic field strength is a critical parameter affecting both the intensity and spectral characteristics of the observable radio emission. Different galactic environments with varying magnetic field strengths may thus exhibit significantly different radio signatures of AGN feedback. Future work should integrate more advanced magnetic field models, preferably using MHD simulations that self-consistently evolve the magnetic field in conjunction with the gas dynamics.

\subsubsection{Particle acceleration efficiency}
\label{sec:particle_acceleration_discussion}
The efficiency of particle acceleration at shocks remains a significant uncertainty in our model. We adopt standard values from the literature for parameters such as injection efficiency \citep[$\eta \sim 10^{-3}$-$10^{-4}$,][]{drury89,berezhko94,kang95} which is controlled by injection parameter $x_\text{inj}$ and the maximum energy ratio \citep[$\xi_\text{max} = 0.05$,][]{keshet03,pfrommer08}, but these values may vary with shock properties and local conditions. We also calculated with different injection parameters, $x_\text{inj}=3.2$ (corresponding to $\eta \sim 10^{-2}$-$10^{-3}$) and $x_\text{inj}=3.9$ (corresponding to $\eta \sim 10^{-4}$-$10^{-5}$). Right panels in Fig.~\ref{fig:beta_comparison} show the results for these different injection parameters. At 1.5 GHz, the total luminosity increases significantly with decreasing $x_\text{inj}$. Specifically, the luminosity for $x_\text{inj}=3.2$ is about 6.3 times higher than that of the fiducial model ($x_\text{inj}=3.6$), and the fiducial model's luminosity is similarly 6.3 times higher than that of $x_\text{inj}=3.9$. Despite these large variations in luminosity, the spectral index $\alpha$ remains nearly constant, suggesting that changes in $x_\text{inj}$ primarily affect the overall energy budget of accelerated electrons without significantly altering their energy distribution. This trend is also reflected in the surface brightness, which follows a similar scaling with $x_\text{inj}$. Future research should more thoroughly explore the parameter space and potentially incorporate advanced models of particle acceleration that account for factors such as magnetic field orientation and pre-existing turbulence.

\subsubsection{Observational systematics}
For weaker shocks, the synchrotron emission may become too faint to be detectable, especially in scenarios where the plasma beta parameter ($\beta$) is large or the injection parameter ($x_\text{inj}$) is high. As shown in our simulations, larger $\beta$ values correspond to weaker magnetic fields, which reduce the synchrotron luminosity. Similarly, higher $x_\text{inj}$ values decrease the injection efficiency of nonthermal electrons, further suppressing the emission, as discussed in Section~\ref{sec:magnetic_field_discussion} and~\ref{sec:particle_acceleration_discussion}. These factors combined can result in radio signals that fall below the sensitivity thresholds of current and next-generation radio telescopes. However, in cases where strong wind events occur, the resulting shocks are capable of producing significantly brighter synchrotron emission. Our simulations demonstrate that such events can generate radio luminosities within the detection capabilities of facilities like the SKA, FAST Core Array, and ngVLA\@. These observations will be crucial for probing the physical conditions of AGN feedback and the properties of the surrounding medium.

While a detailed analysis of observational systematics such as uv-coverage, calibration, and confusion is beyond the scope of this theoretical study, we note that under normal observing conditions, the detectability of the predicted radio emission is feasible with next-generation facilities. For example, at distances of $50-100\,\mathrm{Mpc}$, the angular size of the brightest shock structures corresponds to approximately $5''-9''$, which is well within the resolving capabilities of instruments like SKA-Mid and the FAST Core Array. These facilities are expected to achieve sufficient sensitivity and resolution to detect both the integrated flux and the surface brightness of the extended structures discussed in this work.

\subsubsection{Three-dimensional effects}
Although computationally efficient, our use of 2D axisymmetric simulations necessarily excludes certain 3D effects that could impact shock development and the resulting synchrotron emission. These effects include non-axisymmetric instabilities, turbulent cascades, and complex magnetic field geometries \citep{zhang25}. Full 3D simulations, though computationally demanding, would offer a more comprehensive understanding of shock dynamics and emission properties.

\subsubsection{Diffusion and transport processes}
Our current model assumes that nonthermal electrons remain tightly coupled to the fluid elements in which they are accelerated, neglecting spatial diffusion and other transport processes. In reality, high-energy electrons can diffuse along magnetic field lines and undergo drift motions, potentially altering the spatial distribution of synchrotron emission \citep{osmanov10,chen20}. Including these processes would require solving the cosmic ray transport equation in conjunction with the hydrodynamic equations, a computationally demanding task that would provide more accurate predictions for the emission morphology.

\subsubsection{Thermal bremsstrahlung contamination}
An important observational challenge in interpreting radio emission from wind-driven shocks is the potential contamination from thermal bremsstrahlung (free-free) emission. At radio frequencies, distinguishing between synchrotron emission from shock-accelerated nonthermal electrons and bremsstrahlung emission from the hot thermal plasma can be challenging, particularly when both components have similar spatial distributions and comparable intensities.

\begin{figure}[htbp]
    \centering
    \includegraphics[width=\columnwidth]{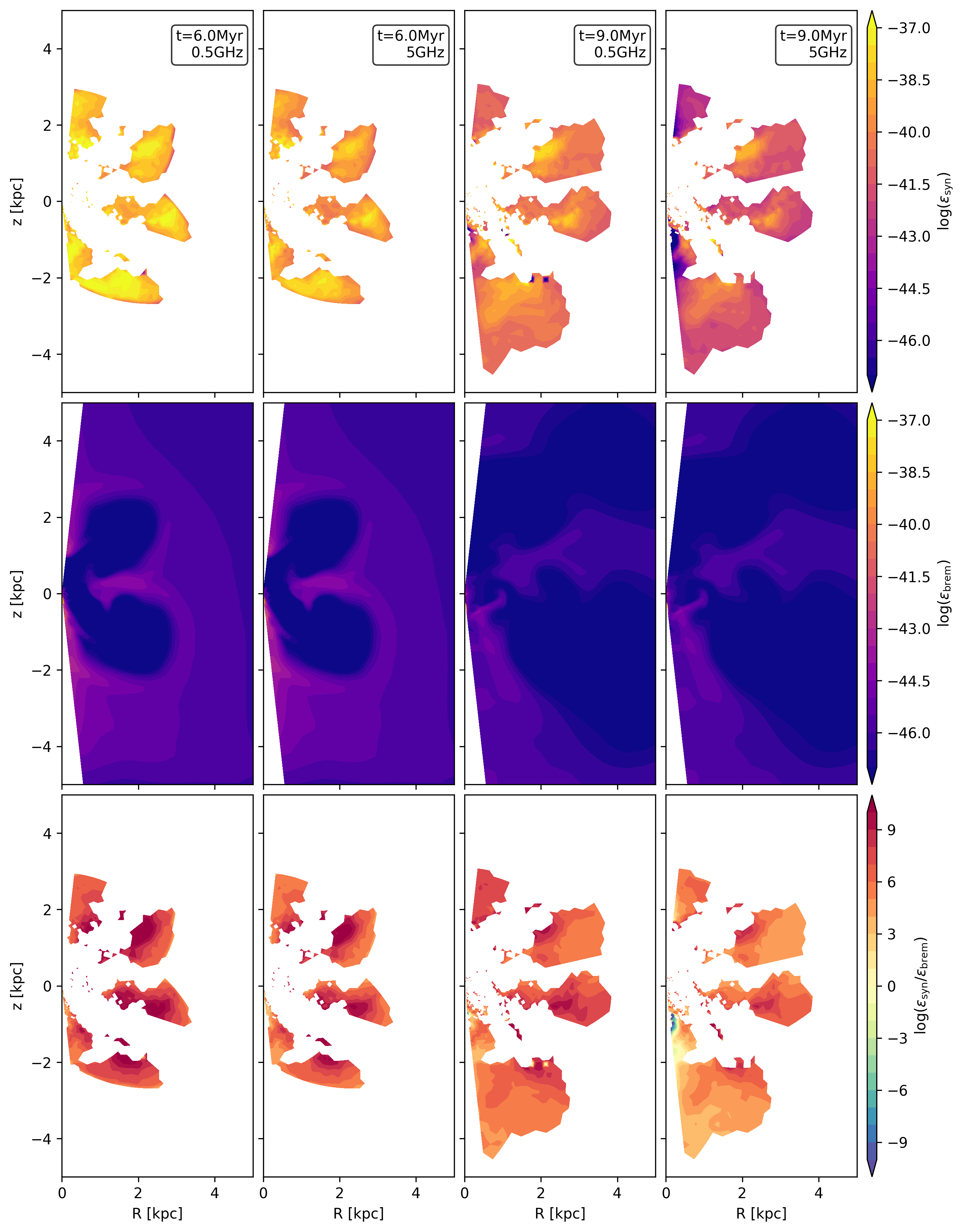}
    \caption{Synchrotron vs.\ thermal bremsstrahlung luminosities in \texttt{fiducial model}. Top panel shows synchrotron emission at $0.5$ and $5.0\,\mathrm{GHz}$ ($t = 6.0$ and $9.0\,\mathrm{Myr}$). Middle panel shows thermal bremsstrahlung at same frequencies. Bottom panel shows synchrotron/bremsstrahlung ratios.}
    \label{fig:synchrotron_ff_comparison}
\end{figure}

To assess this potential contamination, we calculate the thermal bremsstrahlung emission from the hot gas in our simulations using standard formulas \citep{Rybicki79}. The thermal radio emission depends on the electron density and temperature of the gas, with the emissivity scaling as $j_{\nu,\text{ff}} \propto n_e^2 T^{-1/2} \exp(-h\nu/kT)$ at radio frequencies where $h\nu \ll kT$. Fig.~\ref{fig:synchrotron_ff_comparison} compares the predicted synchrotron and thermal bremsstrahlung emission at representative frequencies.

Our calculations show that synchrotron emission typically dominates the radio luminosity at frequencies below approximately $5\,\mathrm{GHz}$ during the peak shock phases, when nonthermal electron acceleration is most efficient. However, thermal bremsstrahlung can become comparable to or even exceed synchrotron emission at higher frequencies when the shock-accelerated electron population cools significantly. This frequency-dependent behavior provides a potential observational discriminant: synchrotron-dominated emission should exhibit steeper spectral indices, while bremsstrahlung-dominated emission typically shows flatter spectra.

\subsection{Broader implications for AGN feedback and galaxy evolution}
\subsubsection{Tracing AGN activity history}
Synchrotron emission from wind-driven shocks offers a unique probe of past AGN activity, potentially uncovering episodes of enhanced accretion that took place millions of years ago. This "fossil record" is especially valuable for understanding the duty cycle and variability of AGN feedback, which operates on timescales far longer than human observation. Our simulations of M32 illustrate how current radio observations can constrain the properties of AGN wind events that occurred about $0.04\,\mathrm{Myr}$ ago, providing insights into the recent history of the central black hole.

\subsubsection{Feedback efficiency across galaxy masses}
The comparison between \texttt{fiducial model} and \texttt{M32 model} reveals notable differences in the scale, intensity, and duration of AGN feedback effects. In massive ellipticals, the stronger gravitational potential and higher black hole mass result in more energetic winds that generate stronger shocks, extending to larger radii ($\sim 5\,\mathrm{kpc}$) and lasting for longer periods (several $\mathrm{Myr}$). In contrast, the shocks in \texttt{M32} are confined to the central $10\,\mathrm{pc}$ and dissipate within around $0.04\,\mathrm{Myr}$.

These differences have significant implications for the overall efficiency of AGN feedback in regulating galaxy evolution. In massive ellipticals, the extended reach and longer duration of shock heating could effectively suppress cooling flows and star formation across much of the galaxy. In dwarf ellipticals like M32, the more confined extent and shorter duration of shock heating imply that AGN feedback may be effective only in the very central regions, with stellar feedback potentially playing a more dominant role at larger radii.

\section{Conclusions}\label{sec:conclusion}
In this study, we investigate the synchrotron emission signatures of AGN-wind-driven shocks in elliptical galaxies through numerical simulations and predict their observational manifestations using current and future radio telescopes. Our main conclusions are as follows:

1. Our fiducial model for a massive elliptical galaxy demonstrates that shocks driven by AGN winds can efficiently accelerate electrons, generating detectable synchrotron radiation. We predict that this emission can reach luminosities of $\sim 10^{29}\,\mathrm{erg\,s^{-1}\,Hz^{-1}}$ in the radio band ($0.5-5\,\mathrm{GHz}$), making it a viable target for next-generation facilities like the SKA, the FAST Core Array, and the ngVLA\@. The emission exhibits a characteristic evolution in space and spectrum: it begins as compact and bright with a flat spectrum (radio spectral index $\alpha \approx -0.4$ to $-0.6$), and evolves into a more diffuse, large-scale structure with a steeper spectrum ($\alpha \approx -0.8$ to $-1.4$) as the shocks weaken and the electron population ages.

2. For the dwarf elliptical galaxy M32, our simulations show that AGN hot winds can produce detectable, albeit weaker and smaller-scale, radio emission. The predicted morphology and intensity show a strong consistency with the R1 source observed by \cite{peng20}, suggesting that hot-wind-driven shocks are a plausible origin for this radio component.

3. It is important to emphasize the caveats associated with our models. The quantitative forecasts presented here are subject to uncertainties stemming from our simplifying assumptions. In particular, the adoption of a constant plasma-$\beta$, fixed particle injection parameters, and the neglect of nonthermal particle transport could introduce order-of-magnitude uncertainties in the predicted emission strength and timescales. While the qualitative trends are robust, refining these quantitative predictions will require more sophisticated physical models in future work.

These results show important implications for our understanding of AGN feedback mechanisms. Our study indicates that AGN winds influence galaxy gas dynamics not only through mechanical energy input but also by producing observable nonthermal radiation. In particular, the associated radio emission serves as a valuable probe for detecting and characterizing AGN feedback, especially in low-luminosity systems where other signatures may be faint or absent.

\begin{acknowledgements}
    We thank the anonymous referee for the constructive comments which greatly improve the presentation of the paper. H.X. and F.Y. are supported by Natural Science Foundation of China (grants No. 12133008, 12192220, 12192223, and 12361161601) and China Manned Space Project (grant No. CMS-CSST-2021-B02). Z.L. is supported by National Natural Science Foundation of China (grant No. 12225302) and National Key Research and Development Program of China (grant No. 2022YFF0503402), and China Manned Space Program (grant No. CMS-CSST-2025-A10).
\end{acknowledgements}

\bibliographystyle{aa}
\bibliography{ref}
\begin{appendix}
    \section{Shock tube test for shock detection algorithm}\label{app:shock_detection}
\begin{figure}[htbp]
        \centering
        \includegraphics[width=\columnwidth]{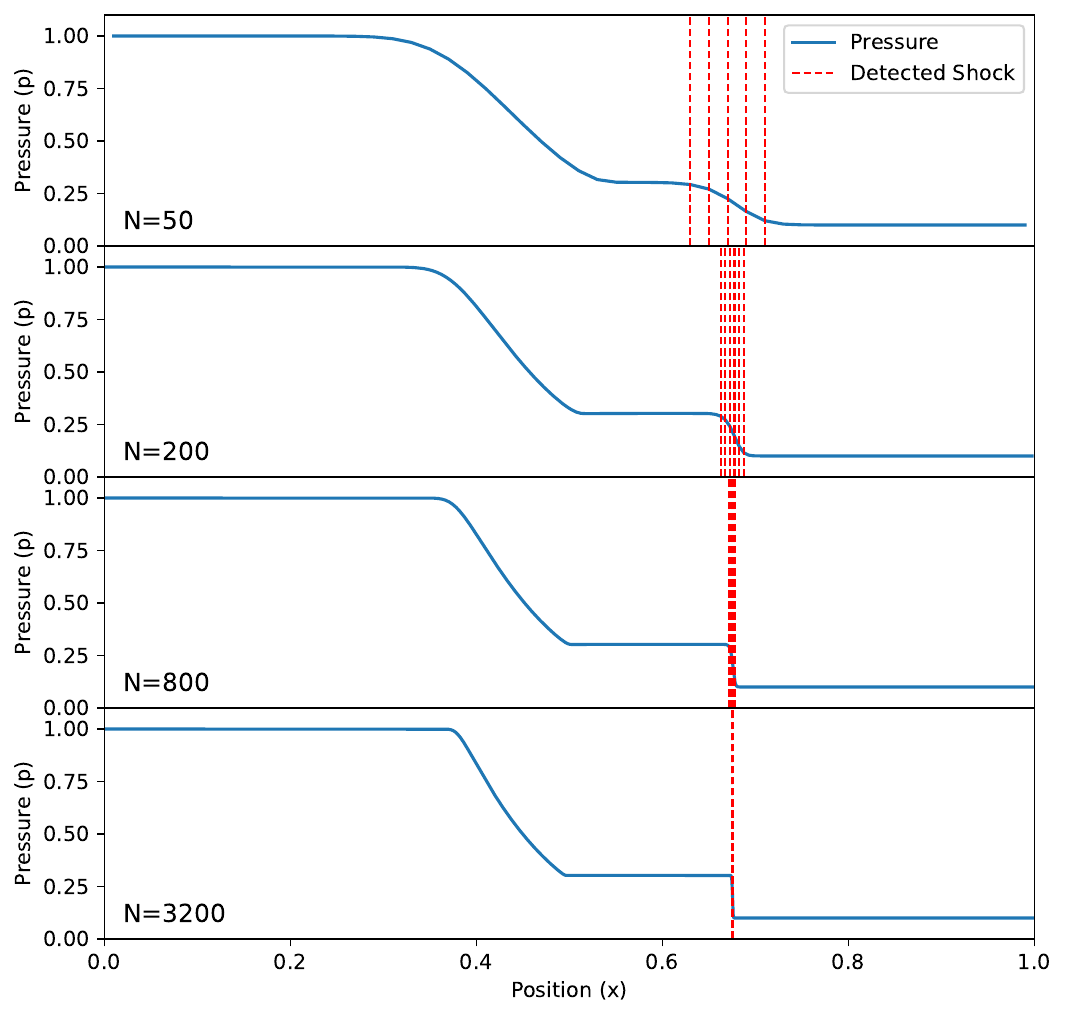}
        \caption{Shock tube test for shock detection algorithm. From top to bottom, the panels show the pressure profile at $t=0.1$ for different grid resolutions ($N=50$, $200$, $800$ and $3200$ cells) in blue solid lines. The red dashed lines indicate the detected shock locations.}
        \label{fig:shock_tube}
    \end{figure}

    To validate our shock detection algorithm, we performed a standard one-dimensional shock tube test \citep{sod78}. The initial conditions of density, velocity, and pressure are given by
    \begin{align}
        (\rho, v, p) = \begin{cases}
                           (1.0, 0.0, 1.0),   & x < 0.5,    \\
                           (0.125, 0.0, 0.1), & x \geq 0.5,
                       \end{cases}
    \end{align}
    And the boundary conditions are set to outflow boundary condition. The adiabatic index is set to $\gamma = 1.4$. We evolved the system using HLLC Riemann solver \citep{toro94} until $t=0.1$ with four sets of uniform grids with $N=50$, $200$, $800$ and $3200$ cells to test the convergence of our shock detection algorithm (see Fig.~\ref{fig:shock_tube}). The shock detection algorithm successfully identified the shock front in all cases, where we set the threshold to $0.05$. As the grid resolution decreases, the shock front becomes less sharp, leading to a slight underestimation of the Mach number. However, even at the lowest resolution ($N=50$), the algorithm still accurately captures the shock location.

    \section{Nonthermal electron acceleration and synchrotron emission}\label{app:particle_synchrotron}

    This appendix provides the detailed mathematical formulation for nonthermal electron acceleration at shocks and their subsequent synchrotron emission, following the approach of \cite{reynolds98,reynolds08,jiang10}. Note that this approach is initially modeled for supernova remnants (SNR). Although the fundamental physics of particle acceleration and radiation in shock waves is common to both SNRs and AGN winds, the differing energy injection mechanisms, environmental complexities, geometries, compositions, and the presence of a strong radiation field must be acknowledged as significant caveats when applying SNR-based models to AGN-driven shocks. These differences can lead to substantial deviations in the predicted spectral evolution and dynamics.

    \subsection{Particle spectrum}\label{app:particle_spectrum}
    The acceleration of particles at shock fronts was extensively investigated in the literature \citep{Blandford87, reynolds08}. The majority of electrons crossing the shock front are assumed to follow a Maxwell-Boltzmann distribution \citep{ep07} as
    \begin{align}
        f_\text{e}(p)=4\pi n_\text{e}\left(\frac{m_\text{e} c^2}{2\pi k T_2}\right)^{3/2}p^2 \exp\left(-\frac{m_\text{e} c^2p^2}{2k T_2}\right),
    \end{align}
    where $n_\text{e}$ is the electron number density, $m_\text{e}$ is the electron mass, $k$ is the Boltzmann constant, $T_2$ is the downstream gas temperature, and $p=p_\text{e}/(m_\text{e} c)$ is dimensionless electron momentum. The momentum threshold for nonthermal electrons is
    \begin{align}
        p_\text{min}=x_\text{inj}\sqrt{\frac{2k T_2}{m_\text{e}c^2}},
    \end{align}
    where $x_\text{inj}\approx 3.3$ to 3.6, a parameter related to the particle injection efficiency $\eta$ defined in Eq.~\eqref{eta_app} \citep{ep07,pfrommer08}. According to the first-order Fermi acceleration mechanism, the accelerated particles follow a power-law momentum distribution as
    \begin{align}
        f_\text{inj}(p)=C p^{-q}\theta(p-p_\text{min}),
    \end{align}
    where $C$ is a normalization constant. The spectral index is given by $q=\frac{\tau+2}{\tau-1}$, where $\tau$ is the shock compression ratio,
    \begin{align}
        \frac{1}{\tau}=\frac{\gamma-1}{\gamma+1}+\frac{2}{\gamma+1}\frac{1}{M_1^2},
    \end{align}
    and $M_1$ is the upstream Mach number.
    The number density of the nonthermal electrons is given by
    \begin{align}
        N_\text{inj}=\int_0^\infty f_\text{inj}(p)\mathrm{d}p=\frac{C p_\text{min}^{1-q}}{q-1}.
    \end{align}
    The corresponding dimensionless injected energy density is
    \begin{align}
        \begin{aligned}
            \mathscr{E}_\text{inj}= & \int_0^\infty f_\text{inj}(p)T(p)\mathrm{d}p                                                                        \\
            =                       & \int_0^\infty f_\text{inj}(p)\left(\sqrt{1+p^2}-1\right)\mathrm{d}p                                                 \\
            =                       & \frac{C}{q-1}\left[\frac{1}{2}\mathscr{B}_\frac{1}{1+p_\text{min}^2}\left(\frac{q-2}{2},\frac{3-q}{2}\right)\right. \\
                                    & \left.+p_\text{min}^{1-q}\left(\sqrt{1+p_\text{min}^2}-1\right)\right],
        \end{aligned}
    \end{align}
    where $T(p)$ is the dimensionless kinetic energy as a function of dimensionless momentum, and $\mathscr{B}_x(a,b)$ is the incomplete beta function. This function is valid only for $2<q<3$, which corresponds to Mach number greater than $\sqrt{5}$. Therefore, weak shocks are excluded. The dimensionless mean kinetic energy is given by
    \begin{align}
        \mathcal{K}_\text{inj}=\frac{p_\text{min}^{q-1}}{2}\mathscr{B}_\frac{1}{1+p_\text{min}^2}\left(\frac{q-2}{2},\frac{3-q}{2}\right)+\sqrt{1+p_\text{min}^2}-1.
    \end{align}
    In the linear regime of CR electron acceleration \citep{ep07, pfrommer08}, the nonthermal distribution is expected to smoothly transition from the thermal distribution, expressed as
    \begin{align}
        f_\text{lin}(p)=f_\text{e}(p_\text{min})\left(\frac{p}{p_\text{min}}\right)^{-q}\theta(p-p_\text{min}).
    \end{align}
    The fraction of electrons accelerated in the linear regime is measured by
    \begin{align}
        \eta_\text{lin}=\frac{\Delta n_\text{lin}}{n_\text{e}}=\frac{4}{\sqrt{\pi}}\frac{x_\text{inj}^3}{q-1}\mathrm{e}^{-x_\text{inj}^2}
        \label{eta_app},
    \end{align}
    with typical values in the range $10^{-3}\sim 10^{-4}$ as inferred from studies of supernova remnants \citep{drury89,berezhko94,kang95} where $\Delta n_\text{lin}$ is the number density of injected electrons in the linear regime. The ratio of nonthermal electron energy density ($\Delta \varepsilon _\text{lin}$) to the increase in downstream thermal energy ($\Delta \varepsilon_\text{ther}$) due to the shock is
    \begin{align}
        \xi_\text{lin}\equiv\frac{\Delta \varepsilon _\text{lin}}{\Delta \varepsilon_\text{ther}}=\frac{\eta_\text{lin}\mathcal{K}_\text{inj}n_\text{e2}m_\text{e}c^2}{(\gamma-1)^{-1}k n_2(T_2-T_1)}.
    \end{align}
    where $n_\text{e2}$ and $n_2$ are the downstream electron number density and total number density respectively, and $T_1$ is the upstream gas temperature.
    The maximum allowed ratio is taken to be $\xi_\text{max}=0.05$ \citep{keshet03,pfrommer08}. Accordingly, the normalization coefficient is corrected as $C=\left(1-\mathrm{e}^{-\delta}\right)\delta^{-1}f_\text{e}(p_\text{min})p_\text{min}^{q}$, where $\delta=\xi_\text{lin}/\xi_\text{max}$.

    \subsection{Synchrotron emission of accelerated electrons}\label{app:synchro}
    Here we primarily follow the spectral evolution of \cite{reynolds98,jiang10}. The energy spectrum of nonthermal electrons at the moment the shock waves first pass through is expressed as
    \begin{align}
        N(\Gamma_0)=f_\text{inj}(p)\frac{\mathrm{d}p}{\mathrm{d}\Gamma_0},
    \end{align}
    where $p=\sqrt{\Gamma_0^2-1}$, $\Gamma$ is the Lorentz factor of the nonthermal electrons, and 0 marks the initial moment when the shock waves first pass through.
    We assume that these nonthermal electrons remain frozen in the fluid element, which continues to move with the fluid flow.
    The energy of an individual electron is dissipated through radiation processes, including synchrotron radiation and inverse Compton scattering, and adiabatic expansion.
    The rate of energy dissipation due to radiation is given by
    \begin{align}
        -\dot{E}_\text{rad}=\frac{4}{3}\sigma_\text{T}c\left(\frac{E}{m_\text{e}c^2}\right)^2\left(\frac{B_\text{eff}^2}{8\pi}\right),
    \end{align}
    where $B_\text{eff}=\sqrt{B^2+B_\text{CMB}^2}$ is the effective magnetic field, and $B_\text{CMB}=3.27\,\mathrm{\mu G}$ is the magnetic field with energy density equivalent to that of the cosmic microwave background radiation at the redshift $z=0$.
    In terms of magnetic field $B$, we assume the ratio between the gas thermal pressure and magnetic pressure, i.e., $\beta\equiv \frac{P_\text{gas}}{B^2/(8\pi)}\approx 10$ \citep{churazov08, jiang10, wang20, wang21}.
    For adiabatic losses, assuming ultra-relativistic electrons with an adiabatic index of $4/3$, the energy density scales as $u_\text{e}\propto V^{-4/3}$, where $V$ is the volume of the fluid element, while the energy of an individual electron scales as $E\propto V^{-1/3}$. This yields
    \begin{align}
        -\dot{E}_\text{ad}=\frac{1}{3}\frac{E}{V}\frac{\mathrm{d}V}{\mathrm{d}t}.
    \end{align}
    Defining $\alpha \equiv \rho/\rho_{2,0}$. Then the above expression can be rewritten as
    \begin{align}
        \dot{E}_\text{ad}=\frac{E}{3\alpha}\frac{\mathrm{d}\alpha}{\mathrm{d}t}.
    \end{align}
    The total dimensionless energy loss rate is then given by
    \begin{align}
        -\frac{\mathrm{d}\Gamma}{\mathrm{d}t}=a B_\text{eff}^2\Gamma^2-\frac{\Gamma}{3\alpha}\frac{\mathrm{d}\alpha}{\mathrm{d}t},
    \end{align}
    where $a=(4e^4)/(9m_\text{e}^3c^5)$ and $e$ is the elementary charge.
    The solution of the above equation is
    \begin{align}
        \Gamma=\alpha^{1/3}\frac{\Gamma_0}{1+A \Gamma_0},\text{where}\  A=a\int_{t_0}^t B_\text{eff}^2\alpha^{1/3}\mathrm{d}t.
        \label{Gamma_app}
    \end{align}
    Also,
    \begin{align}
        \Gamma_0=\frac{\Gamma}{\alpha^{1/3}-A\Gamma}, \quad \frac{\mathrm{d}\Gamma_0}{\mathrm{d}\Gamma}=\alpha^{1/3}\frac{\Gamma_0^2}{\Gamma^2}.
    \end{align}
    By conservation of the number of particles, we have
    \begin{align}
        N(\Gamma)=N(\Gamma_0)\frac{\mathrm{d}\Gamma_0}{\mathrm{d}\Gamma}=N(\Gamma_0)\alpha^{1/3}\frac{\Gamma_0^2}{\Gamma^2}.
        \label{Energy_spectrum_app}
    \end{align}
    For an individual electron, the synchrotron emission is given by
    \begin{align}
        P(\nu)=\frac{\sqrt{3}e^3B}{m_\text{e}c^2}\frac{\nu}{\nu_c}\int_{\nu/\nu_c}^\infty\mathscr{K}_{5/3}(x)\mathrm{d}x,
    \end{align}
    where $\mathscr{K}_{5/3}(x)$is a modified Bessel function of the second kind, and $\nu_c=3e B\Gamma^2/(4\pi m_\text{e}c)$ is the critical frequency.
    It is evident that analytical fits to the synchrotron functions \citep{Fouka13} accelerate the calculation process.
    The synchrotron radiation power per unit volume per unit frequency is given by
    \begin{align}
        J_\nu=\int_{\Gamma_\text{min}}^{\Gamma_\text{max}}P_\nu N(\Gamma)\mathrm{d}\Gamma,
        \label{Synchrotron_emission_app}
    \end{align}
    where the minimum injected electron energy is $\Gamma_\text{0,min}=\sqrt{p_\text{min}^2+1}$ and corresponding $\Gamma$ is given by Eq.~\eqref{Gamma_app}.
    For the maximum injected electron energy, $\Gamma_\text{0,max}$, the timescale of shock wave acceleration and electron cooling must be considered \citep{drury83,reynolds98} as
    \begin{align}
        \begin{aligned}
            t_\text{acc}= & \frac{3}{U_1-U_2}\left(\frac{\kappa_1}{U_1}+\frac{\kappa_2}{U_2}\right)         \\
            =             & \frac{3}{U_1-U_2}\frac{\zeta E c}{3e B}\left(\frac{1}{U_1}+\frac{1}{U_2}\right) \\
            =             & \frac{\zeta E c}{e B U_1^2}\frac{\tau(1+\tau)}{\tau-1},
        \end{aligned}
    \end{align}
    \begin{align}
        t_\text{cool}=\frac{\Gamma_0}{\mathrm{d}\Gamma_0/\mathrm{d}t}=\frac{\Gamma_0}{a\Gamma_0^2B^2}=\frac{1}{a\Gamma_0 B^2},
    \end{align}
    where $\zeta=1$ for strong shock waves and $E = \Gamma_0 m_\text{e}c^2$ is the energy of the electron.
    The maximum energy is achieved when these timescales are equal, which can be expressed as
    \begin{align}
        \Gamma_\text{0,max}=\sqrt{\frac{e U_1^2}{a\zeta m_\text{e}c^2B}\frac{\tau-1}{\tau(\tau+1)}}=\sqrt{\frac{9(m_\text{e}c)^2}{4\zeta e^3}\frac{U_1^2}{B}\frac{\tau-1}{\tau(\tau+1)}}.
        \label{Gamma_max_app}
    \end{align}
\end{appendix}
\end{document}